\documentclass[11pt,a4paper,british]{article}

\newcommand{\blind}{0}

\usepackage[T1]{fontenc}
\usepackage[latin9]{inputenc}
\usepackage{lmodern}
\usepackage{graphicx}
\usepackage{amsthm}
\usepackage{babel}
\usepackage{bm}
\usepackage{epstopdf}
\usepackage{authblk,bibentry,natbib}
\usepackage{mathrsfs}
\usepackage[titletoc,title]{appendix}
\usepackage{amsmath}
\usepackage{amssymb}
\usepackage{setspace}
\usepackage{float}
\usepackage{subcaption}
\usepackage{hhline}
\usepackage{booktabs}
\usepackage{mathtools}
\usepackage{hyperref}
\usepackage{cleveref}
\usepackage{nameref}
\usepackage{thmtools}
\usepackage{pdflscape}
\usepackage[stable]{footmisc}
\usepackage{relsize}
\usepackage{enumitem}
\usepackage{tabularx}
\usepackage{bbm}
\usepackage{datetime}

\usepackage{geometry}
\PassOptionsToPackage{normalem}{ulem}
\usepackage{ulem}

\usepackage{color}
\definecolor{webbrown}{rgb}{0.65, 0.16, 0.16}
\definecolor{RoyalBlue}{rgb}{0.0, 0.14, 0.4}
\definecolor{webgreen}{rgb}{0.0, 0.5, 0.0}
\hypersetup{
colorlinks=true, linktocpage=true, pdfstartpage=1, pdfstartview=FitV,
breaklinks=true, pdfpagemode=UseNone, pageanchor=true, pdfpagemode=UseOutlines,
plainpages=false, bookmarksnumbered, bookmarksopen=true, bookmarksopenlevel=1,
hypertexnames=true, pdfhighlight=/O, urlcolor=webbrown, linkcolor=RoyalBlue, citecolor=webgreen}

\renewcommand{\theequation}{\thesection.\arabic{equation}}
\numberwithin{equation}{section}

\newtheorem*{assumption*}{Assumption}
\crefname{equation}{equation}{equations}

\newcommand*{\doi}[1]{\url{https://doi.org/#1}}

\newcommand{\double}{0}

\newcommand{\finalv}{0}

\if1\double
    
\fi

\begin{document}

\title{Predictive Claim Scores for Dynamic Multi-Product Risk Classification in Insurance}
\if0\blind
    \author{\large Robert Matthijs Verschuren\thanks{
    \if1\double
        \protect\linespread{1.5}\protect\selectfont 
    \fi
    Corresponding author: Amsterdam School of Economics, University of Amsterdam, Roetersstraat 11, 1018 WB, Amsterdam, The Netherlands. E-mail: \href{mailto:r.m.verschuren@uva.nl}
    {\tt r.m.verschuren@uva.nl}}}
    \affil{\it Amsterdam School of Economics, University of Amsterdam}
\fi
\if1\blind
    \author{}
\fi
\if0\finalv
    \date{This version is released on \usdate\today.}
\fi
\if1\finalv
    \date{}
\fi
\maketitle

\begin{abstract}

\noindent It has become standard practice in the non-life insurance industry to employ Generalized Linear Models (GLMs) for insurance pricing. However, these GLMs traditionally work only with \textit{a priori} characteristics of policyholders, while nowadays we increasingly have \textit{a posteriori} information of individual customers available, sometimes even across multiple product categories. In this paper, we therefore consider a dynamic claim score to capture this \textit{a posteriori} information over several product lines. More specifically, we extend the Bonus-Malus-panel model of \citet{boucher2014} and \citet{boucher2018} to include claim scores from other product categories and to allow for non-linear effects of these scores. The application of the resulting multi-product framework to a Dutch property and casualty insurance portfolio shows that the claims experience of individual customers can have a significant impact on the risk classification and that it can be very profitable to account for it.

\end{abstract}

\newpage % Abstract
\section{Introduction} \label{Section1}

It has become the industry standard in non-life insurance to adopt Generalized Linear Models (GLMs) for determining the premium rate structure. Traditionally, these rate structures are based only on \textit{a priori} characteristics of policyholders and do not account for any information available \textit{a posteriori}. In addition, customers often hold multiple policies across different product categories, while insurers tend to focus on policies in a single line of business when designing their premia. However, a lot of individual heterogeneity is typically unaccounted for in these \textit{a priori} univariate rate structures, which may (partially) be captured by information observed \textit{a posteriori} and from other product lines.

Several methods have been introduced in the literature to account for this form of heterogeneity. Common shocks, copulas and Vector GLMs, for instance, can induce a correlation structure between claims from different product categories (see, e.g., \citet{yee2003,bermudez2011,shi2014}). Multivariate random effects and multivariate credibility can additionally accommodate a dynamic correction of the \textit{a priori} rate structure by absorbing any variation not already accounted for by the covariates in GLMs (see, e.g., \citet{englund2008,englund2009,barseghyan2018,pechon2018}). However, \citet{lemaire1998} argues that past claiming behavior is one of the most important determinants of future claim counts and that a Bonus-Malus System (BMS) is therefore more intuitive for this correction. In contrast to the random effect and credibility models, the timing of past claims is now explicitly accounted for in these systems through a claim score as a special case of a Markov process with a finite number of states \citep{kaas2008}. As such, BMSs pose a commercially attractive form of experience rating where only the current level of the score matters instead of the entire claims history.

Despite the appealing framework, these predictive claim scores have primarily been used for designing rate structures for a single product and in a cross-sectional setting. Many authors consider BMSs for car or motor insurance, for instance, to adjust the given static premium and without accounting for any information from other product categories (see, e.g., \citet{pinquet1997,denuit2007,tzougas2014}). A more dynamic approach is followed by \citet{boucher2014}, who argue that it is no longer consistent to use this two-step approach in case of panel or longitudinal data and suggest to estimate the \textit{a posteriori} rate structure in a single step. \citet{boucher2018} further develop the resulting BMS-panel model and, for practical reasons, consider linear effects for the levels of the claim score.

While the BMS-panel model deals with past claiming behavior in a longitudinal set-up, it has thus far only focused on linear effects and claim scores for a single product. This paper therefore extends this BMS-panel model by allowing the claim scores to affect the rate structures of other product lines and by incorporating a natural cubic spline for their effects using a Generalized Additive Model (GAM). In addition, a piecewise linear simplification of the cubic spline is considered in this framework to accommodate an interpretable rate structure in practice with more flexibility than the pure linear specification. This, in turn, allows us to account for information observed \textit{a posteriori} and from other product lines in our rate structures, to identify the cross-selling potential of customers and to investigate the relation between past claiming behavior across different product categories.

The remainder of this paper is organised as follows. In \Cref{Section2}, we briefly highlight the concepts behind the industry standard of GLMs, describe the novel extension of the BMS-panel model and additionally discuss how to determine the optimal claim score. While \Cref{Section3} describes the Dutch property and casualty insurance data set and comments on the exact optimization procedure, we apply this methodology in \Cref{Section4} and elaborate on the results. The final section concludes this paper with a discussion of the most important findings and implications. % 1 Introduction
\section{Modeling framework} \label{Section2}

\subsection{Static \textit{a priori} risk classification} \label{Section2.1}

Among non-life insurance companies, there has been a long tradition of adopting statistical techniques to construct their \textit{a priori} rate structure. These companies are typically interested in predicting the total claim amount $L$ relative to the exposure to risk $e$ in the form of a risk premium. Technically, this risk premium $\pi$ is defined as
\begin{equation}\label{E2.1}
    \pi = \mathbb{E}\left[ \frac{L}{e} \right] = \mathbb{E}\left[ \frac{N}{e} \times \mathbb{E}\left[ \frac{L}{N} \bigg| N > 0 \right] \right] = \mathbb{E}\left[ F \right] \times \mathbb{E}\left[ S \right],
\end{equation}
with $N$, $S = L / N$ and $F = N / e$ the number of claims, the severity of each claim and the claim frequency, respectively, and where independence is assumed between the claim frequency and severity \citep{antonio2012,henckaerts2019}. This assumption can be relaxed by allowing the claim frequencies and severities to interact, but in general it is common practice to model these two components independently (see, e.g., \citet{czado2012,garrido2016}).

Insurers now require predictive models for both the frequency and the severity component to properly estimate these risk premia. In general, they adopt the framework of Generalized Linear Models, where the response variable $Y$ is modelled indirectly through a given link function $g(\cdot)$ as a linear function of explanatory variables $X$ \citep{nelder1972}. More specifically, let $Y_{i,t}$ denote the observation of subject $i$ in period $t$ and let these variables be independently distributed for each subject and period according to some distribution from the exponential family. The mean predictor $\eta_{i,t}$ in a GLM is then given by
\begin{equation}\label{E2.2}
    \eta_{i,t} = g\left( \mu_{i,t} \right) = X_{i,t}^{\prime} \beta \quad \text{for} \quad i = 1, \dots, M, \quad t = 1, \dots, T_i,
\end{equation}
with $\mu_{i,t} = \mathbb{E}\left[ Y_{i,t} | X_{i,t} \right]$ the conditional expectation of $Y_{i,t}$ and $\beta$ a parameter vector containing the risk factors. The vector of covariates $X_{i,t}$ consists of the observable risk characteristics of subject $i$ in period $t$, such as age, gender and additional coverages in the context of non-life insurance, and may contain an element of one to include a constant in the model. In practice, we typically assume a Poisson or Negative Binomial distribution for the claim counts and a Gamma or Inverse-Gaussian distribution for the claim sizes, while a logarithmic link function is commonly adopted to accommodate a multiplicative rate structure \citep{haberman1996}.

While these GLMs have become the industry standard over the last decades, they lead to a static form of risk classification that only takes \textit{a priori} information of policyholders into account. However, the longitudinal set-up in this paper allows us to easily incorporate any \textit{a posteriori} information in the mean predictor to account for past claiming behavior. By additionally including the claims experience from other product categories, we obtain a framework for dynamic multi-product risk classification.

\subsection{Dynamic \textit{a posteriori} risk classification} \label{Section2.2}

While independence is assumed between both subjects and periods in the cross-sectional model, we can account for dependencies between periods in the longitudinal setting. This, in turn, allows us to explicitly incorporate past claiming behavior and dynamically classify risks in our insurance portfolio. In the BMS-panel model of \citet{boucher2014}, past claiming behavior is summarized by a single claim score. This predictive claim score for subject $i$ in period $t+1$ is defined recursively as
\begin{equation}\label{E2.3}
    \ell_{i,t+1} = \min \left( \max \left( \ell_{i,t} + e_{i,t} \mathbbm{1}\left( N_{i,t} = 0 \right) - \Psi \frac{N_{i,t}}{e_{i, t}}, 1 \right), s \right)
\end{equation}
with initial value $\ell_{i,0} = \ell_{0}$ and indicator function $\mathbbm{1}\left( N_{i,t} = 0 \right)$ that equals one for a period without any claims and zero otherwise. The parameters $\Psi$, $s$ and $\ell_{0}$ of this claim score denote a jump parameter, the maximum level of the score and the initial score for new policyholders without any experience yet, respectively. The lowest level of the score as well as the jump after a claim-free period are both fixed at one, since we can already capture their effect indirectly through the parameters $\Psi$ and $s$. We additionally introduce the exposure of risk $e_{i,t}$ into this claim score to account for policies with exposures of less than an entire year, or for policyholders joining or leaving the insurer throughout the year. With this continuous claim score, policyholders who claim more frequently will receive lower scores, whereas policyholders who claim less often will receive higher scores. The level of the score $\ell_{i,t}$ is therefore an indication of the \textit{a posteriori} risk in a policy, since a score of $1$ ($s$) represents policyholders with the highest (lowest) amount of risk. In the BMS literature, this type of score is generally referred to as a system with transition rules $+1/-\Psi$, entry level $\ell_{0}$ and maximum level $s$.

With this claim score, we can now directly incorporate past claiming behavior as an explicit covariate into the linear predictor of a longitudinal GLM. The intuition behind this is that we consider the claim score as a relevant predictor for future claiming behavior, rather than an \textit{ex-post} punishment and reward system for claims in the past. We can even extend this concept for multiple products owned simultaneously by the same policyholder by feeding their claim scores as additional covariates into the predictor as well. If we let superscript $(c)$ denote the product category, then the linear predictor of this multi-product claim score model is given by
\begin{equation}\label{E2.4}
    \eta_{i,t}^{(c)} = g^{(c)}\left(\mu_{i,t}^{(c)}\right) = X_{i,t}^{(c)\prime} \beta^{(c)} + \sum_{j = 1}^{C} f^{(c)}_{j}\left( \ell_{i,t}^{(j)} \right) \quad \text{for} \quad c = 1, \dots, C.
\end{equation}
Here, $f^{(c)}_{j}(\cdot)$ represents some function and the quantity $\exp\left( f^{(c)}_{c}\left( \ell_{i,t}^{(c)} \right) \right)$ is typically called the relativity of the claim score or BMS in case of a logarithmic link function. By additionally requiring that $f^{(c)}_{j}\left(\ell_{i,t}^{(c)}\right) = 0$ whenever $\ell_{i,t}^{(c)} = \ell_{0}$ or unknown, we can account for policyholders without any \textit{a posteriori} information (yet) and for customers holding only a subset of all available products. In turn, the \textit{a priori} risk premia are fully determined by the policyholder's risk characteristics and any effects of the past claiming behavior, if any, are multiplicative to these premia.

While \citet{boucher2014} and \citet{boucher2018} consider linear relativities, or a logarithmic specification for $f(\cdot)$, the transformations $f^{(c)}_{j}(\cdot)$ can be taken in a much more general way. The set of natural cubic splines typically used in GAMs, for instance, can already capture a non-linear as well as a linear effect of the claim score on the response variable \citep{hastie1986}. It can additionally be shown that these splines are optimal among all twice continuously differentiable functions when minimizing the penalized deviance and that they can easily be constructed by a linear combination of so-called B-splines \citep{hastie2009,ohlsson2010}. More importantly, we can express the linear predictor of the multi-product claim score model as a GAM with this set of splines, where, given the claim score specification, parameters can be estimated straightforwardly by maximum likelihood and with standard statistical software for GAMs (see \ref{AppendixB}). However, these cubic splines can lead to complicated non-linear rate structures that are difficult to explain and interpret, so a piecewise linearly segmented rate structure is preferable from a practical perspective. We therefore adopt both natural cubic and linear splines for the functions $f^{(c)}_{j}(\cdot)$ in this paper to allow for non-linear effects of the claim score and to benefit from the existing framework for GAMs. In turn, the resulting multi-product claim score model allows us to dynamically classify the risk profile of policyholders based on their experience in multiple product categories.

\subsection{Optimality of rate structure} \label{Section2.3}

With the multi-product claim score model, we can formulate a rate structure based on the \textit{a priori} characteristics of the policyholders and their past claiming behavior across product categories. Depending on our choice for the claim score parameters $(\Psi, s, \ell_{0})$, different premium rates will result from the estimated model. Moreover, since a lot of different parameter combinations, and thus rate structures, are possible in this framework, a criterion is required to assess their performance. While typical statistical goodness-of-fit measures such as the Akaike Information Criterion (AIC) and Bayesian Information Criterion (BIC) are based on maximized likelihoods, we are more interested in the discriminatory power of the predicted premium from a practitioner's point of view. Our aim in this context is to best identify and distinguish between risky customers and safe customers. A well-known approach for assessing the discriminatory power is based on the Lorenz curve and the Gini index.

The Lorenz curve has first been introduced by \citet{lorenz1905} in the field of welfare economics as a statistical tool to compare two distributions. In case of perfect alignment between the two distributions, the Lorenz curve reduces to the 45-degree line, or the line of equality. Similarly, the greater the discrepancy between the two distributions, the further the Lorenz curve is away from this line of equality. The Gini index is defined as twice the distance between this Lorenz curve and the line of equality, and thus represents a measure of inequality \citep{gini1912}. More importantly, in the context of insurance ratemaking the Lorenz curve and the corresponding Gini index can also be adopted as a measure of risk discrimination (see, e.g., \citet{frees2014}; \citet{henckaerts2019}). To find the Lorenz curve in practice, we can use the following three steps:
{\it\begin{enumerate}[label = \roman{*}), itemsep = 0pt]
    \item Construct the relativity $R_{j} = P_{j}^{\mathrm{A}} / P_{j}^{\mathrm{B}}$ for each policy $j = 1, \dots, H$, where $P_{j}^{\mathrm{B}}$ denotes the risk premium of a benchmark model and $P_{j}^{\mathrm{A}}$ the risk premium of an alternative model;
    \item Order the policies by the relativities $R_{j}$ from lowest to highest;
    \item Calculate
    \begin{equation}\label{E2.8}
        \left( \hat{F}_{P}(\omega), \hat{F}_{L}(\omega) \right) = \left( \frac{\sum_{j = 1}^{H} P_{j}^{\mathrm{B}} \times \mathbbm{1}\left\{F_{H}(R_{j}) \leq \omega\right\}}{\sum_{j = 1}^{H} P_{j}^{\mathrm{B}}}, \frac{\sum_{j = 1}^{H} L_{j} \times \mathbbm{1}\left\{F_{H}(R_{j}) \leq \omega\right\}}{\sum_{j = 1}^{H} L_{j}} \right)
    \end{equation} 
    as a function of $\omega \in [0, 1]$, where $L_{j}$ denotes the actual observed claim amount of policy $j$ and $F_{H}(\cdot)$ the empirical cumulative distribution function of the relativities $R_{j}$.
\end{enumerate}}
\noindent In turn, this ordered Lorenz curve $\left\{\left( \hat{F}_P(\omega) , \hat{F}_L(\omega) \right): \omega \in [0, 1] \right\}$ leads to the empirical Gini index, given by the expression
\begin{equation}\label{E2.9}
    \widehat{\mathrm{Gini}} = 1 - \sum_{j = 0}^{H - 1} \left( \hat{F}_{P}\left(F_{H}(R_{j+1}) \right) - \hat{F}_{P}\left(F_{H}(R_{j}) \right) \right) \left( \hat{F}_{L}\left(F_{H}(R_{j+1}) \right) + \hat{F}_{L}\left(F_{H}(R_{j}) \right) \right)
\end{equation}
from the trapezoidal rule where $R_{0} = 0$ and its asymptotic covariance matrix can be consistently estimated as
\begin{equation}\label{E2.10}
    \hat{\Sigma}_{\mathrm{Gini}} = \frac{4}{\hat{\mu}_{L}^2 \hat{\mu}_{P}^2} \left( 4 \hat{\Sigma}_{h} + \frac{\hat{\mu}_{h}^2}{\hat{\mu}_{L}^2} \hat{\Sigma}_{L} + \frac{\hat{\mu}_{h}^2}{\hat{\mu}_{P}^2} \hat{\Sigma}_{P} - \frac{4 \hat{\mu}_{h}}{\hat{\mu}_{L}} \hat{\Sigma}_{hL} - \frac{4 \hat{\mu}_{h}}{\hat{\mu}_{P}} \hat{\Sigma}_{hP} + \frac{2 \hat{\mu}_{h}^2}{\hat{\mu}_{L} \hat{\mu}_{P}} \hat{\Sigma}_{LP} \right)
\end{equation}
with $h_{j} = \frac{1}{2} (\mu_{L} P^{\mathrm{B}}_{j} F_{L}(R_{j}) + L_{j} \mu_{P} [ 1 - F_{P}(R_{j})])$ for $j = 1, \dots, H$, and using moment-based estimators for all the means and covariances of $L$, $P$ and $h$ \citep{frees2011}. 

Depending on our choice for the benchmark model, there are two versions of the Gini index, namely the simple Gini index and the ratio Gini index. If we simply assume a constant premium for every policy without any risk discrimination, or $P_{j}^{\mathrm{B}} = 1$, we are calculating the simple Lorenz curve and the total degree of risk discrimination of the alternative model. However, we often have an existing framework or benchmark premium rate in place, such as a standard GLM, that we would like to improve. In these cases, it makes more sense to not calculate the total degree of risk discrimination but to compare the risk classification resulting from the alternative model to that from this benchmark model. \citet{frees2014} describe a mini-max strategy for determining which model leads to the best risk classification. They calculate the ratio Gini coefficient for every combination of alternative and benchmark model and select the benchmark model that minimizes the maximal coefficient. The intuition behind this is that the model that minimizes the maximal Gini coefficient is the least vulnerable to alternative specifications. The use of this ratio Gini index has an additional practical advantage, since we can directly relate it to the profit potential of the alternative rate structure over the benchmark structure. If we take $P_{j}^{\mathrm{B}}$ to be the risk premia of a benchmark GLM, then the ratio Gini coefficient (divided by two) quantifies how much more profitable it is, on average, to use, for instance, the multi-product claim score model due to the different ordering of risks. In other words, the ratio Gini index enables us to identify which design of the claim score is most profitable and optimal in the sense of risk discrimination as opposed to the benchmark GLM in non-life insurance. % 2 Modeling framework
\section{Data and empirical considerations} \label{Section3}

\subsection{Property and casualty insurance} \label{Section3.1}

To illustrate the implications of the multi-product claim score model in practice, we apply this model framework to real-world non-life insurance claims. More specifically, we analyze property and casualty insurance data from a sizeable Dutch insurance portfolio containing policies for general liability insurance, home contents insurance, home insurance and travel insurance on a policyholder-specific level. For each of these insurance products, we consider a different set of explanatory variables that are known to be used to construct premium rates in the Netherlands. For more details on the exact covariates used for each product category in this paper, see \ref{AppendixA}.

\begin{figure}[b!]%[ht!]
    \centering
	\begin{tabular}{c c}
        \centering
        \includegraphics[width=0.475\textwidth]{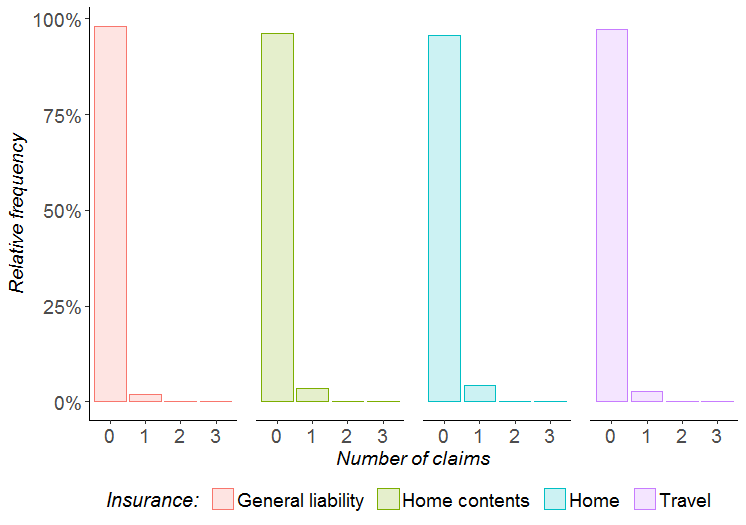}&
        \includegraphics[width=0.475\textwidth]{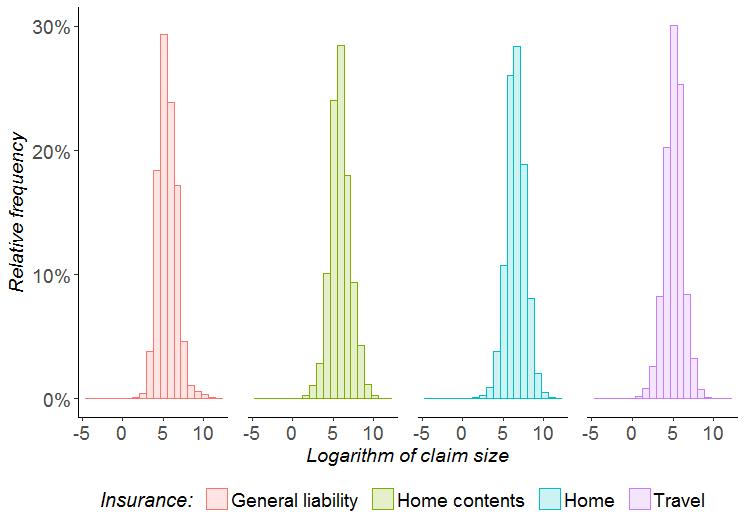}
    \end{tabular}%\vspace{-5pt}
    \caption{Empirical distribution of claim counts (left) and sizes (right) in general liability insurance, home contents insurance, home insurance and travel insurance.}
	\label{F3.1}
\end{figure}

While this Dutch insurance portfolio covers the period of 2012 up to and including 2018, its policies generally have a duration of one year and need to be renewed annually. In addition, policyholders may enter or leave the portfolio at any moment. To cast these policies into a longitudinal framework, we therefore aggregate policies within the same calendar year for each customer and every product category if the policyholder's risk characteristics have remained the same. As a result, we obtain $183{,}690$ observations on general liability insurance in the portfolio, $264{,}348$ observations on home contents insurance, $111{,}018$ observations on home insurance and $363{,}573$ observations on travel insurance. The individual claim counts and sizes for each product category are given in \Cref{F3.1}, where we can clearly see excess zeros in the number of claims and the skewed shape of the claim severities. Note that, for illustrative purposes, the claim severities are shown on a logarithmic scale and that the excess zeros in the claim counts seem to indicate that a Negative Binomial (NB) or Zero-Inflated distribution is more appropriate than a Poisson distribution. More importantly, in \Cref{T3.1} we show how many policyholders also own other products and in \Cref{T3.2} how many claims have occurred in each product line. For general liability insurance, for instance, we observe $34{,}407$ customers, of which $5{,}660$ $[16.45\%]$ own only the general liability insurance product, $13{,}285$ $[38.61\%]$ exactly two products, $11{,}898$ $[34.58\%]$ exactly three products and $3{,}564$ $[10.36\%]$ all four products. Out of the $3{,}520$ general liability claims in total, $511$ $[14.52\%]$ are filed by customers holding only the general liability insurance product, $1{,}192$ $[33.86\%]$ by those holding exactly two products, $1{,}290$ $[36.65\%]$ by those holding exactly three products and $527$ $[14.97\%]$ by those holding all four products. Moreover, in $339$, $147$ and $12$ cases the customers holding exactly two, three and four products, respectively, have also filed at least one claim on the other product(s). In other words, we see that quite a lot of customers have general liability insurance, home contents insurance and/or home insurance, but that there is relatively little overlap between travel insurance and the other three categories. In addition, given that a customer files a claim, we observe a slight yet non-negligible tendency of customers holding exactly two products to have claims in both product categories, while this diminishes as the customer owns more products.

%\clearpage

\begin{table}[t!]%[ht!]
    \caption{Number of policyholders and corresponding percentages in square brackets in each insurance product category.}
    \label{T3.1}\vspace{-3pt}
	\centerline{\scalebox{0.80}{\begin{tabular}{l r@{ }r@{}l r@{ }r@{}l r@{ }r@{}l r@{ }r@{}l r@{ }r@{}l r@{ }r@{}l }
		\toprule \addlinespace[1ex] \vspace{1pt}
		& \multicolumn{12}{c}{\textbf{Products owned}} & \\
		\cline{2-13} \addlinespace[0.4ex]
		\textbf{Policyholders} & \multicolumn{3}{c}{\textbf{1}} & \multicolumn{3}{c}{\textbf{2}} & \multicolumn{3}{c}{\textbf{3}} & \multicolumn{3}{c}{\textbf{4}} & \multicolumn{3}{c}{\textbf{Total}} \\ \hline \addlinespace[0.4ex]
		\textit{General liability} & 5{,}660& [16.45\%]& & 13{,}285& [38.61\%]& & 11{,}898& [34.58\%]& & 3{,}564& [10.36\%]& & 34{,}407& [100.00\%]& \\
		\textit{Home contents} & 10{,}272& [22.88\%]& & 18{,}333& [40.84\%]& & 12{,}724& [28.34\%]& & 3{,}564& [7.94\%]& & 44{,}893& [100.00\%]& \\
		\textit{Home} & 688& [3.44\%]& & 5{,}612& [28.08\%]& & 10{,}119& [50.64\%]& & 3{,}564& [17.84\%]& & 19{,}983& [100.00\%]& \\
		\textit{Travel} & 73{,}412& [89.01\%]& & 2{,}042& [2.48\%]& & 3{,}461& [4.20\%]& & 3{,}564& [4.32\%]& & 82{,}479& [100.00\%]& \\
		\bottomrule
	\end{tabular}}}
\end{table} %\newpage

\begin{table}[t!]%[ht!]
    \caption{Number of claims in each insurance product category with in parentheses taking into account that at least one claim has also been filed in the other categories and corresponding percentages in square brackets.}
    \label{T3.2}\vspace{-3pt}
	\centerline{\scalebox{0.80}{\begin{tabular}{l r@{ }r@{}l r@{ }r@{}l r@{ }r@{}l r@{ }r@{}l r@{ }r@{}l r@{ }r@{}l }
		\toprule \addlinespace[1ex] \vspace{1pt}
		& \multicolumn{12}{c}{\textbf{Products owned}} & \\
		\cline{2-13} \addlinespace[0.4ex]
		\textbf{Claims} & \multicolumn{3}{c}{\textbf{1}} & \multicolumn{3}{c}{\textbf{2}} & \multicolumn{3}{c}{\textbf{3}} & \multicolumn{3}{c}{\textbf{4}} & \multicolumn{3}{c}{\textbf{Total}} \\ \hline \addlinespace[0.4ex]
		\textit{General liability} & 511& (0)& & 1{,}192& (339)& & 1{,}290& (147)& & 527& (12)& & 3{,}520& (498)& \\
		& [14.52\%& (0.00\%)&] & [33.86\%& (68.07\%)&] & [36.65\%& (29.52\%)&] & [14.97\%& (2.41\%)&] & [100.00\%& (100.00\%)&] \\
		\textit{Home contents} & 1{,}546& (0)& & 3{,}820& (870)& & 3{,}508& (204)& & 1{,}297& (16)& & 10{,}171& (1{,}090)& \\
		& [15.20\%& (0.00\%)&] & [37.56\%& (79.82\%)&] & [34.49\%& (18.72\%)&] & [12.75\%& (1.47\%)&] & [100.00\%& (100.00\%)&] \\
		\textit{Home} & 63& (0)& & 1{,}285& (522)& & 2{,}506& (174)& & 1{,}127& (16)& & 4{,}981& (712)& \\
		& [1.26\%& (0.00\%)&] & [25.80\%& (73.31\%)&] & [50.31\%& (24.44\%)&] & [22.63\%& (2.25\%)&] & [100.00\%& (100.00\%)&] \\
		\textit{Travel} & 9{,}148& (0)& & 296& (59)& & 645& (35)& & 683& (16)& & 10{,}772& (110)& \\
		& [84.92\%& (0.00\%)&] & [2.75\%& (53.64\%)&] & [5.99\%& (31.82\%)&] & [6.34\%& (14.55\%)&] & [100.00\%& (100.00\%)&] \\
		\bottomrule
	\end{tabular}}}
\end{table} %\newpage

\subsection{Optimization methodology} \label{Section3.2}

Using the Dutch insurance portfolio, we estimate the multi-product claim score model developed in this paper. We model the claim frequencies and severities independently, and assume a Poisson and NB distribution for the claim counts and a Gamma and Inverse-Gaussian (IG) distribution for the claim sizes, both with logarithmic link function. Moreover, we consider an ordinary GLM for the claim severities and apply the multi-product framework to the claim frequencies, with explanatory variables as given in \ref{AppendixA} for each product category, without any interaction effects. While these assumptions can be relaxed, they are adopted nonetheless since they correspond to standard practices in the non-life insurance industry.

Under these assumptions, we apply the multi-product framework to the property and casualty insurance data. Given the claim score parameters, we estimate this framework by penalized Maximum Likelihood (ML), or Penalized Iteratively Re-weighted Least Squares (PIRLS), which we describe in detail in \ref{AppendixB} and can be performed efficiently in \texttt{R} with the package \texttt{mgcv} developed by \citet{wood2006}. However, rather than letting the smoothing penalty determine the number of parameters for the B-splines, we employ $k = 4$ parameters for all of them to sufficiently account for non-linearities using the $k - 1 = 3$ effective degrees of freedom. We additionally replace the centering constraint on these splines by the \textit{a priori} constraint introduced earlier that $f_{j}\left(\ell_{i,t}\right) = 0$ whenever $\ell_{i,t} = \ell_{0}$ or unknown, equivalent to having no \textit{a posteriori} information. This, in turn, allows us to exploit all information available, both on customers with or without any \textit{a posteriori} information and with or without all products. Finally, we compare the multi-product framework developed in this paper to the case of linear claim score effects similar to the BMS-panel model of \citet{boucher2014} and \citet{boucher2018} to assess the value of our extension.

The above methodology assumes known claim score parameters, while in practice these are unknown as well. We therefore determine the optimal parameters independently for each product by a grid search in terms of the ratio Gini index with a standard GLM as benchmark. More specifically, we estimate the claim score model for each product separately on training data from the period of 2012 up to and including 2017, and select the parameters that lead to the best ratio Gini index for test data from 2018. We additionally impose the restriction that each claim score level, after truncation towards $\ell_{0}$, must contain at least $0.01\%$ of the training set's exposure to avoid parameter combinations that can lead to unobserved levels. While we consider the values $\left\{ 1, 2, \dots, s - 1 \right\}$ for $\Psi$ and $\left\{ 3, 4, \dots, 25 \right\}$ for $s$, we implement the set $\left\{ 2, 3, \dots, s - 1 \right\}$ for $\ell_{0}$ in case policyholders have no prior claiming experience. 

However, in practice policyholders often switch between insurers or have been a customer at the insurer previously, meaning that they do in fact have claiming experience prior to our data. In car or motor insurance, for instance, the number of claim-free years and the level of the claim score are usually known within an insurance company or exchanged between insurers since BMSs are widely implemented in this insurance branch, but this is unfortunately not the case for the product branches that we consider. Nonetheless, for the period of 2005 up to and including 2011, we do have access to the claims history at the insurer, albeit not the individual risk characteristics of the policyholders. As a result, all the claims filed, if any, by $12{,}361$ customers are available for general liability insurance in this period, $13{,}210$ customers for home contents insurance, $5{,}460$ customers for home insurance and $13{,}112$ customers for travel insurance. As a proxy to the unobserved prior claiming experience, we can therefore already construct the claim score and use its level at the end of this seven-year period to initialize the claim score for policyholders that have been a customer at the insurer prior to our data. Given the optimal claim score parameters for each product category, we can then specify and estimate the multi-product framework. As such, the multi-product claim score model remains tractable and allows us to incorporate past claiming behavior across product categories for insurance pricing. % 3 Data description
\section{Applications in non-life insurance} \label{Section4}

\subsection{Static univariate risk profiles} \label{Section4.1}

Based on the Dutch insurance portfolio and the methodology described earlier, we explore how well a standard GLM can describe insurance data. While we adopt the abbreviations in \Cref{T4.1.1} henceforth, we show the resulting out-of-sample error distributions for the estimated GLMs in \Cref{F4.1.1} with parameter estimates reported in \ref{AppendixC}. Moreover, \Cref{T4.1.2} depicts the maximal ratio Gini coefficients for the static GLMs, where we include both the Poisson and NB distribution for the claim frequencies and both the Gamma and IG distribution for the claim severities.

From the prediction errors, it is apparent that the magnitude of the out-of-sample errors differs substantially across product categories. For home contents insurance, for instance, we obtain the largest errors, whereas the predicted premia for general liability insurance appear to be much closer to the realized expenses than for the other product categories. More importantly, we find that the prediction errors are distributed almost the same regardless of the model specification. This is additionally supported by the parameter estimates reported in \ref{AppendixC} which are roughly the same for the two frequency and the two severity components. The distribution underlying the static GLMs therefore does not seem to really affect the prediction errors or to matter that much for the goodness-of-fit.

\begin{table}[t!]%[ht!]
    \caption{Model abbreviations for different combinations of static univariate frequency and severity models.}
    \label{T4.1.1}\vspace{-3pt}
	\centerline{\scalebox{0.80}{\begin{tabular}{l l l }
	    \toprule \addlinespace[1ex] \vspace{1pt}
		\textbf{Abbreviation} & \textbf{Frequency model} & \textbf{Severity model} \\ \hline \addlinespace[0.4ex]
		\textit{GLM-PG} & Poisson GLM & Gamma GLM \\
		\textit{GLM-PIG} & Poisson GLM & Inverse-Gaussian GLM \\
		\textit{GLM-NBG} & Negative Binomial GLM & Gamma GLM \\
		\textit{GLM-NBIG} & Negative Binomial GLM & Inverse-Gaussian GLM \\
		\bottomrule
	\end{tabular}}} \vspace{-1pt}
\end{table} %\newpage

\begin{figure}[t!]%[ht!]
    \centering
	\begin{tabular}{c c}
        \centering
        \begin{subfigure}{0.475\textwidth}
            \centering
            \includegraphics[width=\textwidth]{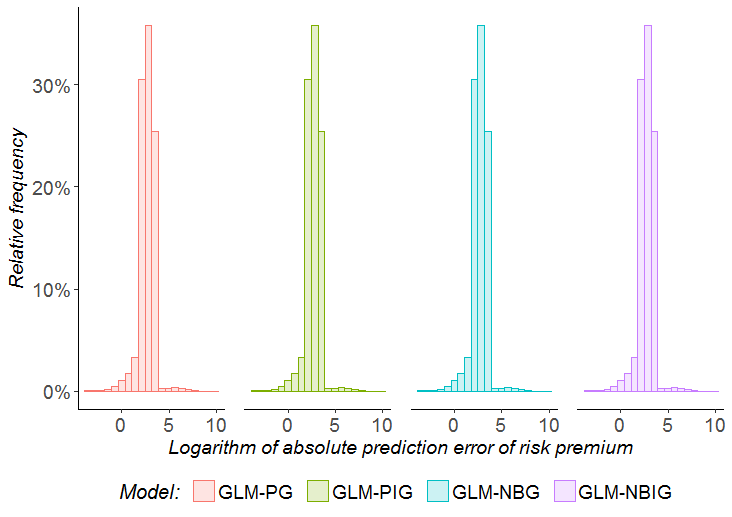}
            \caption{General liability insurance}
            \label{F4.1.1a}
        \end{subfigure}
        \begin{subfigure}{0.475\textwidth}
            \centering
            \includegraphics[width=\textwidth]{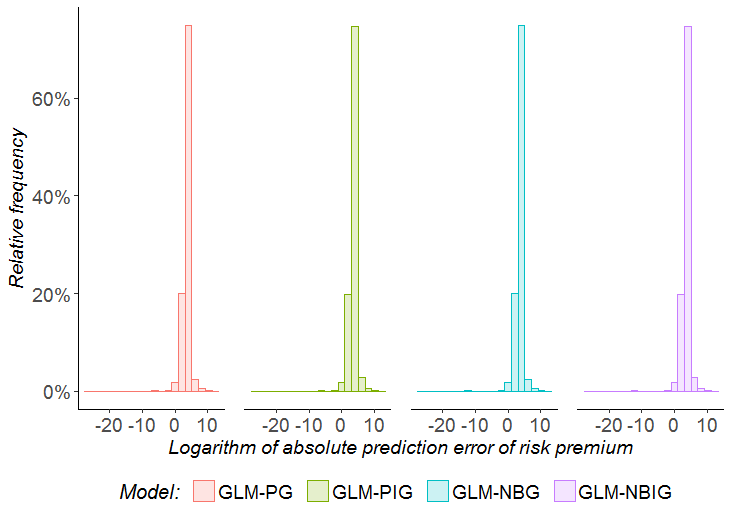}
            \caption{Home contents insurance}
            \label{F4.1.1b}
        \end{subfigure}
    \end{tabular}
    
    \vspace{3pt}
    
    \begin{tabular}{c c}
        \centering
        \begin{subfigure}{0.475\textwidth}
            \centering
            \includegraphics[width=\textwidth]{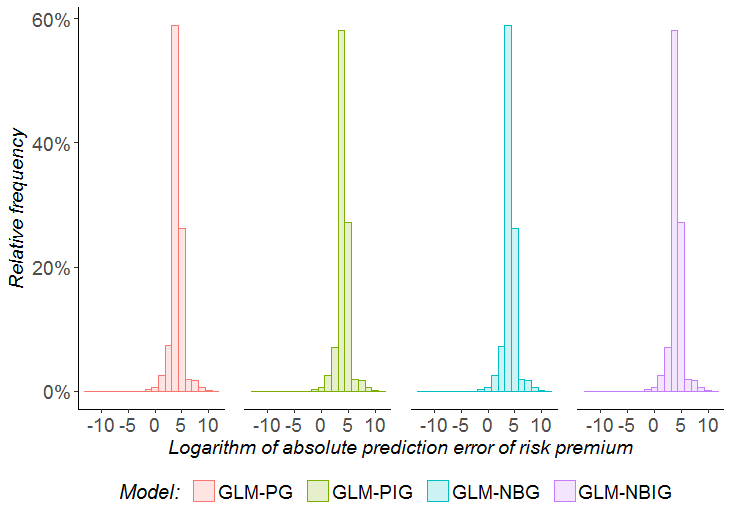}
            \caption{Home insurance}
            \label{F4.1.1c}
        \end{subfigure}
        \begin{subfigure}{0.475\textwidth}
            \centering
            \includegraphics[width=\textwidth]{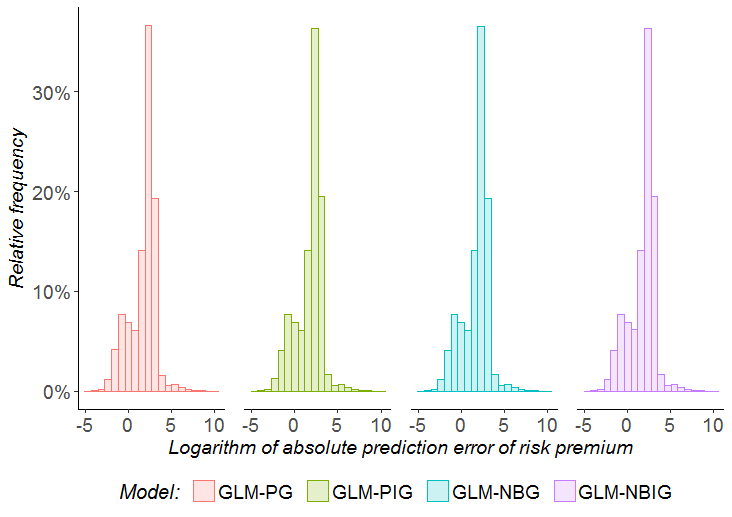}
            \caption{Travel insurance}
            \label{F4.1.1d}
        \end{subfigure}
    \end{tabular}%\vspace{-5pt}
    \caption{Out-of-sample risk premia errors for general liability insurance (panel (a)), home contents insurance (panel (b)), home insurance (panel (c)) and travel insurance (panel (d)) with static univariate risk classification.}
	\label{F4.1.1} \vspace{-1pt}
\end{figure}%\clearpage

\begin{table}[t!]%[ht!]
    \caption{Maximal ratio Gini coefficients in percentages with corresponding standard errors in parenthesis and rank for each product category with static univariate risk classification.}
    \label{T4.1.2}\vspace{-3pt}
	\centerline{\scalebox{0.80}{\begin{tabular}{l r@{ }r@{}l r c r@{ }r@{}l r c r@{ }r@{}l r c r@{ }r@{}l r }
	    \toprule \addlinespace[1ex] \vspace{1pt}
	    & \multicolumn{4}{c}{\textbf{General liability}} &  & \multicolumn{4}{c}{\textbf{Home contents}} &  & \multicolumn{4}{c}{\textbf{Home}} &  & \multicolumn{4}{c}{\textbf{Travel}} \\
		\cline{2-5} \cline{7-10} \cline{12-15} \cline{17-20} \addlinespace[0.4ex]
		\textbf{Benchmark} & \multicolumn{3}{c}{\textbf{Coefficient}} & \textbf{Rank} &  & \multicolumn{3}{c}{\textbf{Coefficient}} & \textbf{Rank} &  & \multicolumn{3}{c}{\textbf{Coefficient}} & \textbf{Rank} &  & \multicolumn{3}{c}{\textbf{Coefficient}} & \textbf{Rank} \\ \hline \addlinespace[0.4ex]
		\textit{GLM-PG} & 10.38& (7.92)& & 3 & & 13.41& (4.77)& & 2 & & 10.31& (7.67)& & 3 & & 15.08& (3.54)& & 4 \\
		\textit{GLM-PIG} & 10.43& (7.87)& & 4 & & 17.66& (4.50)& & 4 & & 2.06& (7.64)& & 2 & & 7.54& (4.03)& & 2 \\
		\textit{GLM-NBG} & \textbf{0.28}& \textbf{(8.77)}& & \textbf{1} & & \textbf{-13.30}& \textbf{(4.73)}& & \textbf{1} & & 10.32& (7.66)& & 4 & & 11.07& (3.21)& & 4 \\
		\textit{GLM-NBIG} & 0.55& (8.59)& & 2 & & 17.35& (4.50)& & 3 & & \textbf{-1.99}& \textbf{(7.60)}& & \textbf{1} & & \textbf{-7.43}& \textbf{(3.92)}& & \textbf{1} \\
		\bottomrule
	\end{tabular}}}
\end{table} %\newpage

However, in terms of the ratio Gini index, we do find a substantial impact of the distribution underlying the static GLMs. Based on the mini-max strategy, we find that the static GLM-NBG is the least vulnerable to alternative model choices for general liability and home contents insurance, while the static GLM-NBIG is the least vulnerable for home and travel insurance. Moreover, these results indicate that the NB distribution can, on average, be improved much less in terms of profit potential than the Poisson distribution for the frequency component. For the severity component of the static GLMs, on the other hand, it depends on the product category at hand whether the Gamma distribution or the IG distribution can, on average, be improved the least in terms of profit potential.

\subsection{Dynamic univariate risk profiles} \label{Section4.2}

\begin{figure}[b!]%[ht!]
    \centering
    \begin{tabular}{c c}
        \centering
        \includegraphics[width=0.45\textwidth]{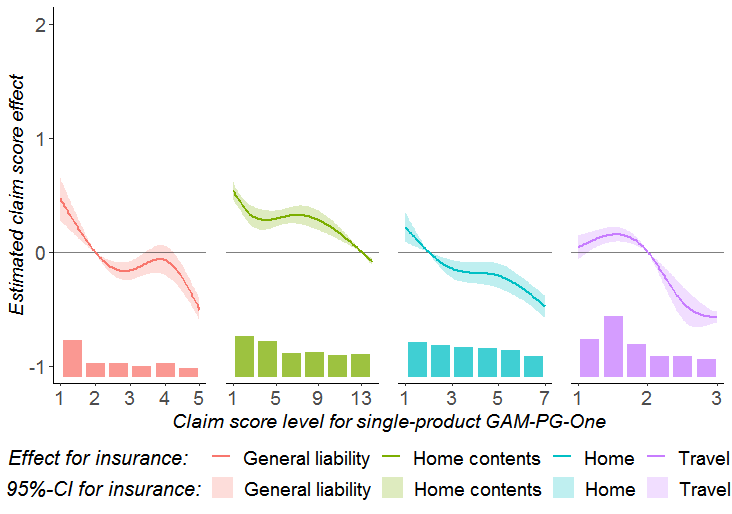}&
        \includegraphics[width=0.45\textwidth]{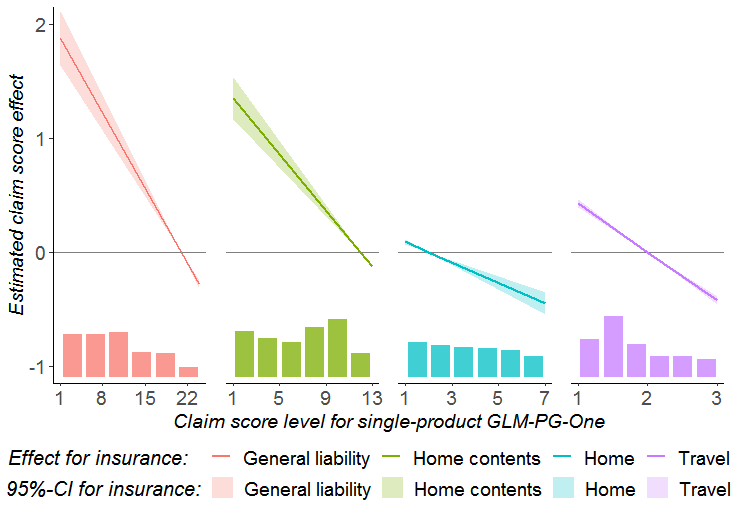}
    \end{tabular}\vspace{-3pt}
    \caption{Estimated claim score effects with corresponding $95\%$ confidence intervals and observed claim frequencies on general liability insurance, home contents insurance, home insurance and travel insurance for single-product GAM-PG-One (left) and GLM-PG-One (right).}
	\label{F4.2.1} %\vspace{-26pt}
\end{figure}

While the standard GLM is a form of static risk classification, we can also create dynamic risk profiles from the claims experience of individual policyholders. Using the claim score introduced earlier we account for this claims experience on a single product and optimize the parameters $(\Psi, s, \ell_{0})$ of the claim scores for each product category separately. The resulting one-product models consider both Poisson and NB distributed claim frequencies, both Gamma and IG distributed claim severities and use either GAM or GLM specifications, and are abbreviated in \Cref{T4.2.1}. Moreover, the optimal set of claim score parameters for each model is given in \Cref{T4.2.2}, whereas we show the effect of these claim scores in \Cref{F4.2.1}. Finally, \Cref{T4.2.3} presents the maximal ratio Gini coefficients when we include these dynamic univariate risk profiles, with all parameter estimates reported in \ref{AppendixC}.

\begin{table}[b!]%[ht!]
    \vspace{-3pt}
    \caption{Model abbreviations for different combinations of dynamic univariate frequency and severity models.}
    \label{T4.2.1}\vspace{-3pt}
	\centerline{\scalebox{0.80}{\begin{tabular}{l l l }
	    \toprule \addlinespace[1ex] \vspace{1pt}
		\textbf{Abbreviation} & \textbf{Frequency model} & \textbf{Severity model} \\ \hline \addlinespace[0.4ex]
		\textit{GAM-PG-One} & Poisson one-product claim score GAM & Gamma GLM \\
		\textit{GAM-PIG-One} & Poisson one-product claim score GAM & Inverse-Gaussian GLM \\
		\textit{GAM-NBG-One} & Negative Binomial one-product claim score GAM & Gamma GLM \\
		\textit{GAM-NBIG-One} & Negative Binomial one-product claim score GAM & Inverse-Gaussian GLM \\
		\textit{GLM-PG-One} & Poisson one-product claim score GLM & Gamma GLM \\
		\textit{GLM-PIG-One} & Poisson one-product claim score GLM & Inverse-Gaussian GLM \\
		\textit{GLM-NBG-One} & Negative Binomial one-product claim score GLM & Gamma GLM \\
		\textit{GLM-NBIG-One} & Negative Binomial one-product claim score GLM & Inverse-Gaussian GLM \\
		\bottomrule
	\end{tabular}}} \vspace{5pt}
\end{table} %\newpage

\begin{table}[b!]%[ht!]
    \caption{Optimal claim score parameters of dynamic univariate frequency models for each product category.}
    \label{T4.2.2}\vspace{-3pt}
	\centerline{\scalebox{0.80}{\begin{tabular}{l r@{ }r@{ }r@{}l r@{ }r@{ }r@{}l r@{ }r@{ }r@{}l r@{ }r@{ }r@{}l }
	    \toprule \addlinespace[1ex] \vspace{1pt}
	    & \multicolumn{16}{c}{\textbf{Optimal claim score parameters $(\Psi, s, \ell_{0})$}} \\
		\cline{2-17} \addlinespace[0.4ex]
		\textbf{Frequency model} & \multicolumn{4}{c}{\textbf{General liability}} & \multicolumn{4}{c}{\textbf{Home contents}} & \multicolumn{4}{c}{\textbf{Home}} & \multicolumn{4}{c}{\textbf{Travel}} \\ \hline \addlinespace[0.4ex]
		\textit{GAM-PG-One} & \textbf{(2,}& \textbf{5,}& \textbf{2)}& & (13,& 14,& 13)& & (1,& 7,& 2)& & (2,& 3,& 2)& \\
		\textit{GAM-PIG-One} & (2,& 5,& 2)& & (13,& 14,& 13)& & (1,& 7,& 2)& & \textbf{(2,}& \textbf{3,}& \textbf{2)}& \\
		\textit{GAM-NBG-One} & (2,& 5,& 2)& & \textbf{(13,}& \textbf{14,}& \textbf{13)}& & (1,& 7,& 2)& & (2,& 3,& 2)& \\
		\textit{GAM-NBIG-One} & (2,& 5,& 2)& & (13,& 14,& 13)& & (1,& 7,& 2)& & (2,& 3,& 2)& \\
		\textit{GLM-PG-One} & (6,& 24,& 21)& & (1,& 13,& 12)& & (1,& 7,& 2)& & (2,& 3,& 2)& \\
		\textit{GLM-PIG-One} & (6,& 24,& 21)& & (1,& 13,& 12)& & \textbf{(1,}& \textbf{7,}& \textbf{2)}& & (2,& 3,& 2)& \\
		\textit{GLM-NBG-One} & (6,& 24,& 21)& & (1,& 13,& 12)& & (1,& 7,& 2)& & (2,& 3,& 2)& \\
		\textit{GLM-NBIG-One} & \hspace{16.3pt} (5,& 20,& 17)& & \hspace{10.2pt} (1,& 13,& 12)& & (1,& 7,& 2)& & (2,& 3,& 2)& \\
		\bottomrule
	\end{tabular}}} \vspace{5pt}
\end{table} %\newpage

 %\newpage%\clearpage

\begin{table}[t!]%[ht!]
    \caption{Maximal ratio Gini coefficients in percentages with corresponding standard errors in parenthesis and rank for each product category with dynamic univariate risk classification.}
    \label{T4.2.3}\vspace{-3pt}
	\centerline{\scalebox{0.80}{\begin{tabular}{l r@{ }r@{}l r c r@{ }r@{}l r c r@{ }r@{}l r c r@{ }r@{}l r }
	    \toprule \addlinespace[1ex] \vspace{1pt}
	    & \multicolumn{4}{c}{\textbf{General liability}} &  & \multicolumn{4}{c}{\textbf{Home contents}} &  & \multicolumn{4}{c}{\textbf{Home}} &  & \multicolumn{4}{c}{\textbf{Travel}} \\
		\cline{2-5} \cline{7-10} \cline{12-15} \cline{17-20} \addlinespace[0.4ex]
		\textbf{Benchmark} & \multicolumn{3}{c}{\textbf{Coefficient}} & \textbf{Rank} &  & \multicolumn{3}{c}{\textbf{Coefficient}} & \textbf{Rank} &  & \multicolumn{3}{c}{\textbf{Coefficient}} & \textbf{Rank} &  & \multicolumn{3}{c}{\textbf{Coefficient}} & \textbf{Rank} \\ \hline \addlinespace[0.4ex]
		\textit{GLM-PG} & 11.12& (7.67)& & 7 &  & 15.45& (4.98)& & 5 &  & 10.66& (7.62)& & 8 &  & 33.12& (4.05)& & 11 \\
		\textit{GLM-PIG} & 11.12& (7.67)& & 6 &  & 24.13& (4.67)& & 12 &  & 8.23& (7.63)& & 4 &  & 32.42& (3.99)& & 9 \\
		\textit{GLM-NBG} & 11.12& (7.67)& & 9 &  & 15.11& (4.97)& & 4 &  & 10.63& (7.62)& & 6 &  & 33.14& (4.04)& & 12 \\
		\textit{GLM-NBIG} & 11.12& (7.67)& & 8 &  & 23.96& (4.67)& & 11 &  & 8.17& (7.63)& & 3 &  & 32.45& (3.98)& & 10 \\ \hline \addlinespace[0.2ex]
		\textit{GAM-PG-One} & \textbf{7.89}& \textbf{(8.04)}& & \textbf{1} &  & 19.93& (4.47)& & 7 &  & 16.88& (7.72)& & 12 &  & 18.48& (3.18)& & 3 \\
		\textit{GAM-PIG-One} & 7.89& (8.04)& & 2 &  & 21.32& (4.44)& & 8 &  & 11.05& (7.52)& & 10 &  & \textbf{17.68}& \textbf{(3.59)}& & \textbf{1} \\
		\textit{GAM-NBG-One} & 7.89& (8.04)& & 3 &  & \textbf{5.93}& \textbf{(4.73)}& & \textbf{1} &  & 16.81& (7.72)& & 11 &  & 18.61& (3.20)& & 4 \\
		\textit{GAM-NBIG-One} & 7.89& (8.04)& & 4 &  & 19.37& (4.43)& & 6 &  & 11.02& (7.52)& & 9 &  & 17.93& (3.60)& & 2 \\
		\textit{GLM-PG-One} & 11.54& (7.68)& & 11 &  & 14.15& (5.06)& & 3 &  & 10.64& (7.69)& & 7 &  & 25.26& (4.14)& & 7 \\
		\textit{GLM-PIG-One} & 11.54& (7.68)& & 12 &  & 22.37& (4.68)& & 10 &  & \textbf{6.98}& \textbf{(7.66)}& & \textbf{1} &  & 25.23& (4.14)& & 6 \\
		\textit{GLM-NBG-One} & 11.42& (7.69)& & 10 &  & 13.96& (5.06)& & 2 &  & 10.62& (7.69)& & 5 &  & 25.29& (4.20)& & 8 \\
		\textit{GLM-NBIG-One} & 10.63& (7.75)& & 5 &  & 22.12& (4.68)& & 9 &  & 7.19& (7.68)& & 2 &  & 25.20& (4.14)& & 5 \\
		\bottomrule
	\end{tabular}}}
\end{table} %\newpage

In \Cref{T4.2.2}, we see that the distribution for the claim frequencies and severities is not very important for the optimal claim score parameters. More specifically, there can be substantial differences between the GLM and GAM specification, but among the one-product GAMs we find that exactly the same parameters are optimal, whereas the one-product GLMs can lead to different but roughly the same parameters. As a consequence, we also obtain almost the same claim score effects for the one-product GAMs and GLMs as those shown for GAM-PG-One and GLM-PG-One in \Cref{F4.2.1}, respectively.

From these claim score effects in \Cref{F4.2.1}, we observe that in general policyholders who claim more frequently are more risky, whereas policyholders who claim less frequently are less risky. However, we do find some exceptions to this rule in case of the one-product GAMs. For general liability insurance, for instance, we find that policyholders with the highest or lowest claim score receive a relatively large discount or surcharge, respectively, but that policyholders with a score between these two extremes receive approximately the same small discount. A similar pattern is observed for home contents insurance, where policyholders with a claim score between the highest and lowest score receive approximately the same surcharge. This non-monotonic relation is partially caused by the relatively small exposure in low claim scores, since the majority of policyholders does not claim and obtains high claim scores, but is primarily inherent from the data. At the bottom of \Cref{F4.2.1}, we see that the cubic claim score effects actually largely follow the non-monotonic relation observed between the claim score and the claim frequency given that score. The one-product GAMs therefore allow us to investigate and identify these non-monotonicities, whereas the one-product GLMs simply assume a single slope coefficient for all claim scores that is only accurate for the scores that we observe most. As such, the linear claim score can essentially be seen as a linear approximation of the cubic claim score, where its effect is mainly determined by the good risks, or the policyholders with high claim scores.

Even though the distribution for the claim frequencies and severities does not seem to be very important for the optimal claim score parameters, it does in fact affect our mini-max strategy. \Cref{T4.2.3} reports, for instance, that the maximal ratio Gini coefficient can be quite different depending on the distributional assumptions, and that NB distributed claim frequencies seem to be slightly less vulnerable to alternative model choices than Poisson distributed claim frequencies. More importantly, \Cref{T4.2.3} shows that for all four product categories the one-product model is an improvement over the regular GLM and that the standard GLM is now, on average, much more vulnerable to alternative model choices. A straightforward Likelihood Ratio (LR) test leads to the same conclusion in-sample for every single case in \Cref{T.C.10} in \ref{AppendixC}. While the one-product GAMs mostly lead to the highest ratio Gini coefficients and therefore seem the most promising, they are only optimal in the sense of our mini-max strategy for general liability, home contents and travel insurance. It actually turns out that the one-product GAMs PG-One, NBG-One and PIG-One are the least vulnerable to alternative model choices for general liability, home contents and travel insurance, respectively, but the one-product GLM PIG-One for home insurance due to near-linear cubic effects. This, in turn, implies that accounting for the claims experience of a single product is a large improvement over the standard GLM, and that the one-product GAMs typically seem to outperform the one-product GLMs.

\subsection{Dynamic multivariate risk profiles} \label{Section4.3}

\begin{figure}%[t!]%[ht!]
    \centering
    \begin{subfigure}{\textwidth}
        \centering
        \begin{tabular}{c c}
            \centering
            \includegraphics[width=0.45\textwidth]{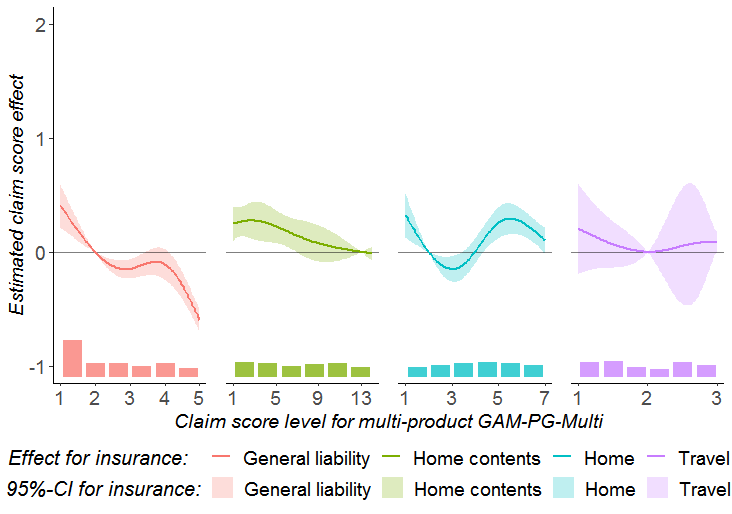}&
            \includegraphics[width=0.45\textwidth]{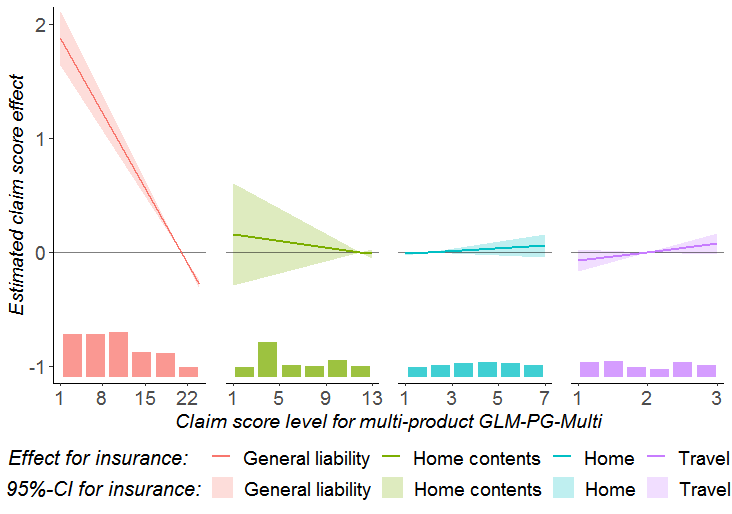}
        \end{tabular}\vspace{-6pt}
        \caption{Effect on general liability insurance}
        \label{F4.3.1a}
    \end{subfigure}
    
    \vspace{3pt}
    
    \begin{subfigure}{\textwidth}
        \centering
        \begin{tabular}{c c}
            \centering
            \includegraphics[width=0.45\textwidth]{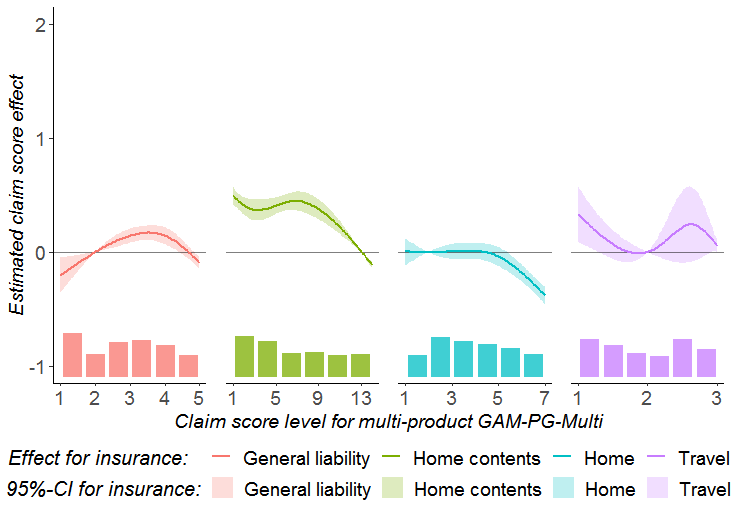}&
            \includegraphics[width=0.45\textwidth]{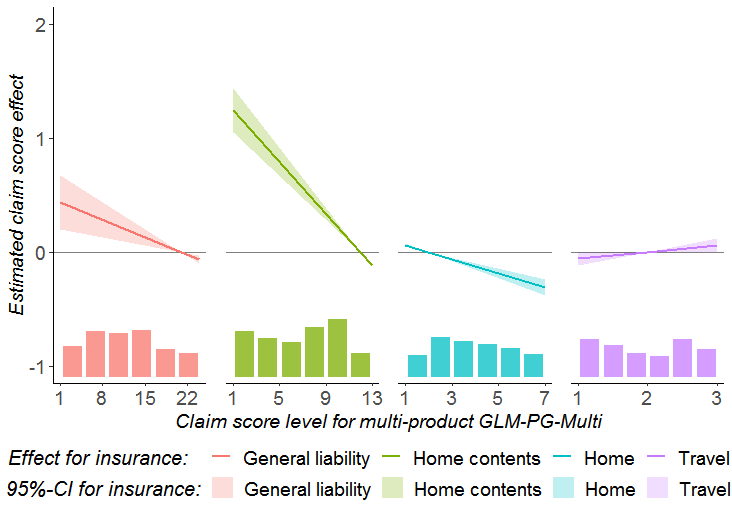}
        \end{tabular}\vspace{-6pt}
        \caption{Effect on home contents insurance}
        \label{F4.3.1b}
    \end{subfigure}
    
    \vspace{3pt}
    
    \begin{subfigure}{\textwidth}
        \centering
        \begin{tabular}{c c}
            \centering
            \includegraphics[width=0.45\textwidth]{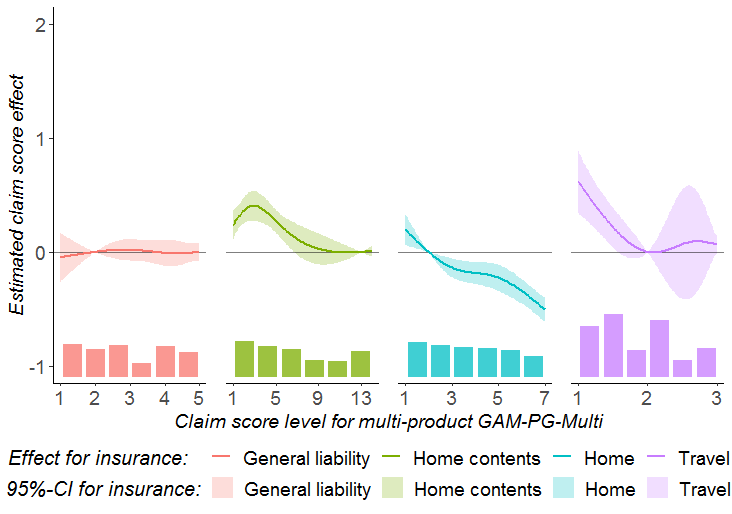}&
            \includegraphics[width=0.45\textwidth]{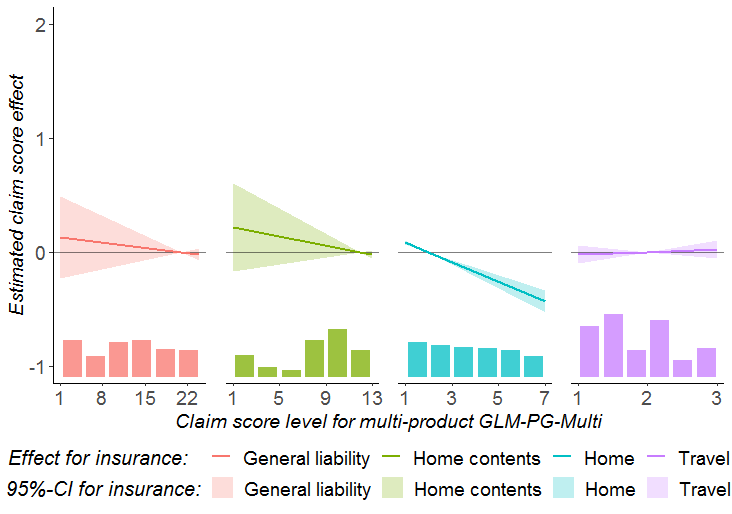}
        \end{tabular}\vspace{-6pt}
        \caption{Effect on home insurance}
        \label{F4.3.1c}
    \end{subfigure}
    
    \vspace{3pt}
    
    \begin{subfigure}{\textwidth}
        \centering
        \begin{tabular}{c c}
            \centering
            \includegraphics[width=0.45\textwidth]{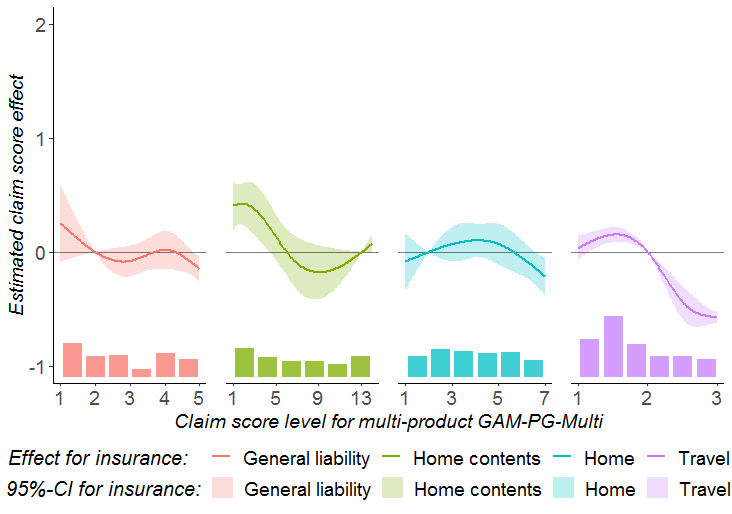}&
            \includegraphics[width=0.45\textwidth]{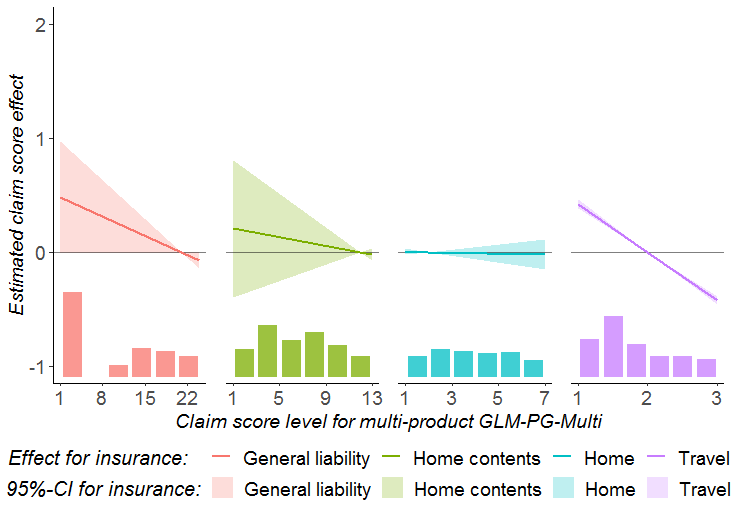}
        \end{tabular}\vspace{-6pt}
        \caption{Effect on travel insurance}
        \label{F4.3.1d}
    \end{subfigure}\vspace{-3pt}
    \caption{Estimated claim score effects with corresponding $95\%$ confidence intervals and observed claim frequencies on general liability insurance (panel (a)), home contents insurance (panel (b)), home insurance (panel (c)) and travel insurance (panel (d)) for multi-product GAM-PG-Multi (left) and GLM-PG-Multi (right).}
	\label{F4.3.1}
\end{figure}

In contrast to the one-product models, we can additionally account for the claims experience across multiple product categories. As such, we extend the one-product models by incorporating the dynamic claim score on the other products of the policyholder, if any, given the previously optimized claim score parameters. We abbreviate the resulting multi-product models in \Cref{T4.3.1}, where we consider both Poisson and NB distributed claim frequencies, both Gamma and IG distributed claim severities and use either GAM or GLM specifications. Moreover, we display the effects of the multi-product claim scores in \Cref{F4.3.1} for each product category separately, where the multi-product GAMs and GLMs again lead to almost the same claim score effects and we therefore only show those for GAM-PG-Multi and GLM-PG-Multi. Finally, \Cref{T4.3.2} presents the maximal ratio Gini coefficients when we include these dynamic multivariate risk profiles, with all parameter estimates reported in \ref{AppendixC}.

From the claim score effects in \Cref{F4.3.1}, we again observe that in general there is a negative relation between the risk of a customer on a certain product and the claim score on that same product. However, this relation is far less clear and appears more complicated for the claim scores on other products. For general liability insurance in \Cref{F4.3.1a}, for instance, we find that policyholders merely possessing home contents insurance are associated with more risk and that this also holds true for all claim scores for travel insurance in case of the multi-product GAMs. This, in turn, implies that insurers should not target customers holding home contents and/or travel insurance with cross-selling offers since we expect these customers to receive low claim scores or claim relatively often on these other products. Note that the relatively large confidence bands for general liability, home contents and travel insurance result from a lack of policyholders with these claim scores, since most policyholders claim very few, more claim score levels lead to sparser distributions and there is relatively little overlap from travel insurance with the other insurance products. The cubic spline approach thus seems more representative of the uncertainty of the claim score effects than the linear approach.

%\vspace{8pt}
\begin{table}[b!]%[ht!]
    \caption{Model abbreviations for different combinations of dynamic multivariate frequency and severity models.}
    \label{T4.3.1}\vspace{-3pt}
	\centerline{\scalebox{0.80}{\begin{tabular}{l l l }
	    \toprule \addlinespace[1ex] \vspace{1pt}
		\textbf{Abbreviation} & \textbf{Frequency model} & \textbf{Severity model} \\ \hline \addlinespace[0.4ex]
		\textit{GAM-PG-Multi} & Poisson multi-product claim score GAM & Gamma GLM \\
		\textit{GAM-PIG-Multi} & Poisson multi-product claim score GAM & Inverse-Gaussian GLM \\
		\textit{GAM-NBG-Multi} & Negative Binomial multi-product claim score GAM & Gamma GLM \\
		\textit{GAM-NBIG-Multi} & Negative Binomial multi-product claim score GAM & Inverse-Gaussian GLM \\
		\textit{GLM-PG-Multi} & Poisson multi-product claim score GLM & Gamma GLM \\
		\textit{GLM-PIG-Multi} & Poisson multi-product claim score GLM & Inverse-Gaussian GLM \\
		\textit{GLM-NBG-Multi} & Negative Binomial multi-product claim score GLM & Gamma GLM \\
		\textit{GLM-NBIG-Multi} & Negative Binomial multi-product claim score GLM & Inverse-Gaussian GLM \\
		\bottomrule
	\end{tabular}}}
\end{table} %\newpage

%%% Multi-product splines %%%
 %\clearpage%\newpage%\clearpage

%%% Ratio Gini coefficients and SE's %%%
\begin{table}[t!]%[ht!]
    \caption{Maximal ratio Gini coefficients in percentages with corresponding standard errors in parenthesis and rank for each product category with dynamic multivariate risk classification.}
    \label{T4.3.2}\vspace{-3pt}
	\centerline{\scalebox{0.80}{\begin{tabular}{l r@{ }r@{}l r c r@{ }r@{}l r c r@{ }r@{}l r c r@{ }r@{}l r }
	    \toprule \addlinespace[1ex] \vspace{1pt}
	    & \multicolumn{4}{c}{\textbf{General liability}} &  & \multicolumn{4}{c}{\textbf{Home contents}} &  & \multicolumn{4}{c}{\textbf{Home}} &  & \multicolumn{4}{c}{\textbf{Travel}} \\
		\cline{2-5} \cline{7-10} \cline{12-15} \cline{17-20} \addlinespace[0.4ex]
		\textbf{Benchmark} & \multicolumn{3}{c}{\textbf{Coefficient}} & \textbf{Rank} &  & \multicolumn{3}{c}{\textbf{Coefficient}} & \textbf{Rank} &  & \multicolumn{3}{c}{\textbf{Coefficient}} & \textbf{Rank} &  & \multicolumn{3}{c}{\textbf{Coefficient}} & \textbf{Rank} \\ \hline \addlinespace[0.4ex]
		\textit{GLM-PG} & 15.96& (7.98)& & 17 &  & 18.16& (4.99)& & 9 &  & 12.38& (7.68)& & 10 &  & 33.12& (4.05)& & 19 \\
		\textit{GLM-PIG} & 15.96& (7.98)& & 16 &  & 24.13& (4.67)& & 20 &  & 10.10& (7.59)& & 8 &  & 32.42& (3.99)& & 17 \\
		\textit{GLM-NBG} & 15.96& (7.98)& & 15 &  & 18.09& (4.99)& & 8 &  & 12.32& (7.67)& & 9 &  & 33.14& (4.04)& & 20 \\
		\textit{GLM-NBIG} & 15.96& (7.98)& & 14 &  & 23.96& (4.67)& & 19 &  & 10.08& (7.59)& & 7 &  & 32.45& (3.98)& & 18 \\ \hline \addlinespace[0.2ex]
		\textit{GAM-PG-One} & 11.29& (7.33)& & 7 &  & 19.93& (4.47)& & 11 &  & 19.24& (7.78)& & 20 &  & 19.13& (3.18)& & 3 \\
		\textit{GAM-PIG-One} & 11.29& (7.33)& & 8 &  & 21.32& (4.44)& & 16 &  & 14.42& (7.62)& & 14 &  & \textbf{18.43}& \textbf{(3.42)}& & \textbf{1} \\
		\textit{GAM-NBG-One} & 11.27& (7.32)& & 5 &  & \textbf{6.25}& \textbf{(4.75)}& & \textbf{1} &  & 19.16& (7.77)& & 19 &  & 19.28& (3.20)& & 4 \\
		\textit{GAM-NBIG-One} & 11.27& (7.32)& & 6 &  & 19.37& (4.43)& & 10 &  & 14.40& (7.62)& & 13 &  & 18.54& (3.49)& & 2 \\
		\textit{GLM-PG-One} & 16.47& (8.04)& & 18 &  & 16.14& (4.92)& & 7 &  & 14.70& (7.76)& & 15 &  & 25.26& (4.14)& & 15 \\
		\textit{GLM-PIG-One} & 16.47& (8.04)& & 19 &  & 22.37& (4.68)& & 18 &  & 8.37& (7.59)& & 4 &  & 25.23& (4.14)& & 14 \\
		\textit{GLM-NBG-One} & 16.60& (8.03)& & 20 &  & 16.07& (4.91)& & 6 &  & 14.74& (7.76)& & 16 &  & 25.29& (4.20)& & 16 \\
		\textit{GLM-NBIG-One} & 15.94& (8.07)& & 13 &  & 22.12& (4.68)& & 17 &  & 8.36& (7.59)& & 3 &  & 25.20& (4.14)& & 13 \\ \hline \addlinespace[0.2ex]
		\textit{GAM-PG-Multi} & 6.56& (8.04)& & 4 &  & 10.35& (4.72)& & 3 &  & 15.28& (7.67)& & 17 &  & 19.97& (3.24)& & 7 \\
		\textit{GAM-PIG-Multi} & 6.56& (8.04)& & 3 &  & 20.51& (4.39)& & 13 &  & 12.68& (7.58)& & 12 &  & 19.47& (3.39)& & 5 \\
		\textit{GAM-NBG-Multi} & \textbf{6.56}& \textbf{(8.04)}& & \textbf{1} &  & 8.45& (4.80)& & 2 &  & 15.28& (7.67)& & 18 &  & 20.11& (3.26)& & 8 \\
		\textit{GAM-NBIG-Multi} & 6.56& (8.04)& & 2 &  & 19.94& (4.41)& & 12 &  & 12.67& (7.58)& & 11 &  & 19.58& (3.41)& & 6 \\
		\textit{GLM-PG-Multi} & 14.44& (7.98)& & 9 &  & 16.04& (4.93)& & 5 &  & 9.88& (7.64)& & 6 &  & 24.78& (4.20)& & 11 \\
		\textit{GLM-PIG-Multi} & 14.44& (7.98)& & 10 &  & 20.90& (4.82)& & 15 &  & 7.12& (7.61)& & 2 &  & 24.17& (4.17)& & 10 \\
		\textit{GLM-NBG-Multi} & 14.47& (7.97)& & 11 &  & 15.94& (4.92)& & 4 &  & 9.86& (7.64)& & 5 &  & 24.78& (4.20)& & 12 \\
		\textit{GLM-NBIG-Multi} & 14.94& (8.04)& & 12 &  & 20.84& (4.81)& & 14 &  & \textbf{6.94}& \textbf{(7.59)}& & \textbf{1} &  & 24.16& (4.15)& & 9 \\
		\bottomrule
	\end{tabular}}}
\end{table} %\newpage

Interestingly enough, the multi-product GLMs do not indicate these subtleties in the claim score due to a lack of exposure in customers owning multiple products and simultaneously having low claim scores. More explicitly, since the effects of the claim scores in these multi-product GLMs are linear, they are essentially based on a weighted average of all the observed claim scores and do not reflect the non-monotonic claim frequencies actually observed in \Cref{F4.3.1}. However, in \Cref{T3.1} and \Cref{T3.2} we see that most policyholders do not claim (at all) and, as a result, end up with high claim scores. The estimates for the linear claim scores are therefore dominated by policyholders with high claim scores and seem a rather poor linear approximation of the cubic claim scores that is primarily appropriate for the good risks. The flexibility of the multi-product GAMs, on the other hand, allows us to adjust for this lack of exposure by employing multiple cubic splines instead of forcing a single linear relation for all claim scores. This, in turn, enables the cubic claim scores to again largely capture the non-monotonicities in the claim frequencies. As such, the effects of the multi-product GLMs seem a result of misspecification, whereas the effects of the multi-product GAMs a data-driven result.

Not surprisingly, the mini-max strategy of the ratio Gini coefficients seems to consistently favor the cubic claim score effects over the linear claim score effects. \Cref{T4.3.2}, for instance, shows that for general liability, home contents and travel insurance the cubic specification is still the least vulnerable to alternative rate structures, while for home insurance the linear specification is the least vulnerable due to near-linear cubic effects. Moreover, the claims experience in other product categories appears only to be useful for risk classification in case of general liability and home insurance. For home contents and travel insurance, we find that this multi-product claims experience is less useful for risk classification and that it is actually more effective to only account for the claims experience in their own product category. Nonetheless, a standard LR test does indicate in \Cref{T.C.15} in \ref{AppendixC} that the multi-product model significantly outperforms the one-product model in-sample for all cubic and some linear specifications and each product category, including home contents and travel insurance.

While this out-of-sample result may seem surprising at first sight, it can primarily be ascribed to two factors. Since most policyholders do not claim (at all) and thus end up with high claim scores, there is a huge excess of zeros in the insurance portfolio, in particular when considering multiple product categories. For home contents insurance, for instance, we observe relatively few policyholders with claims in multiple product categories compared with general liability and home insurance, and we therefore also observe little variation in the claim scores on the other products. In case of travel insurance, we do observe a large pool of policyholders, but few of these customers actually hold multiple insurance products. As a result, there is little information to gain for home contents and travel insurance by accounting for the claims experience in other product categories and it is actually sufficient to merely incorporate the claims experience in their own product category.

\subsection{Piecewise linear simplification} \label{Section4.4}

\begin{figure}[t!]%[ht!]
    \centering
	\begin{tabular}{c c}
        \centering
        \begin{subfigure}{0.45\textwidth}
            \centering
            \includegraphics[width=\textwidth]{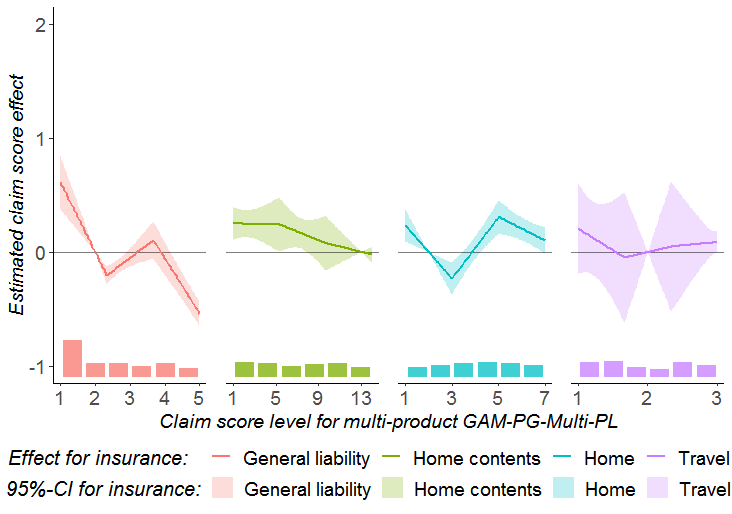}
        \caption{Effect on general liability insurance}
        \label{F4.4.1a}
        \end{subfigure}
        \begin{subfigure}{0.45\textwidth}
            \centering
            \includegraphics[width=\textwidth]{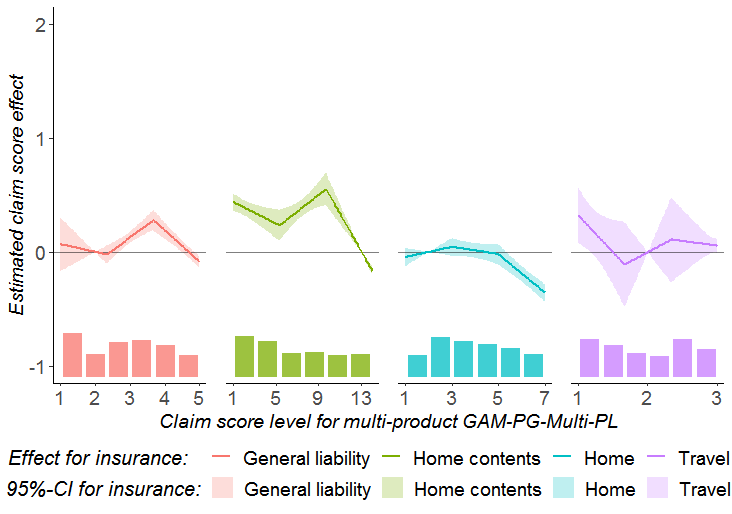}
            \caption{Effect on home contents insurance}
            \label{F4.4.1b}
        \end{subfigure}
    \end{tabular}
    
    \vspace{3pt}
    
    \begin{tabular}{c c}
        \centering
        \begin{subfigure}{0.45\textwidth}
            \centering
            \includegraphics[width=\textwidth]{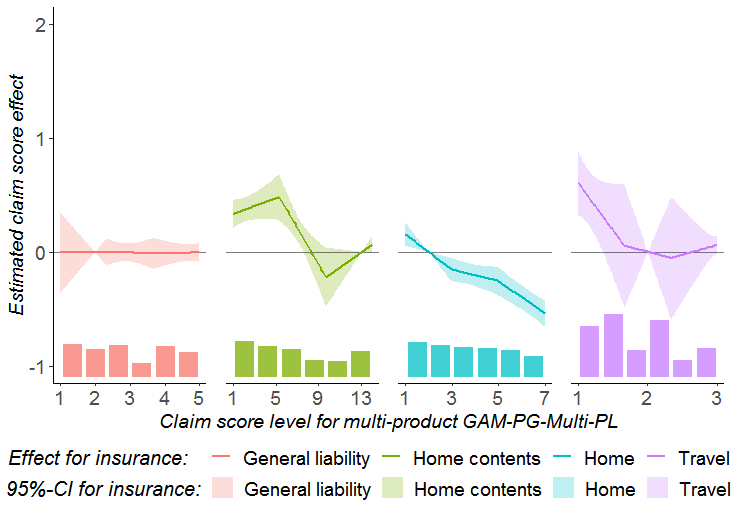}
        \caption{Effect on home insurance}
        \label{F4.4.1c}
        \end{subfigure}
        \begin{subfigure}{0.45\textwidth}
            \centering
            \includegraphics[width=\textwidth]{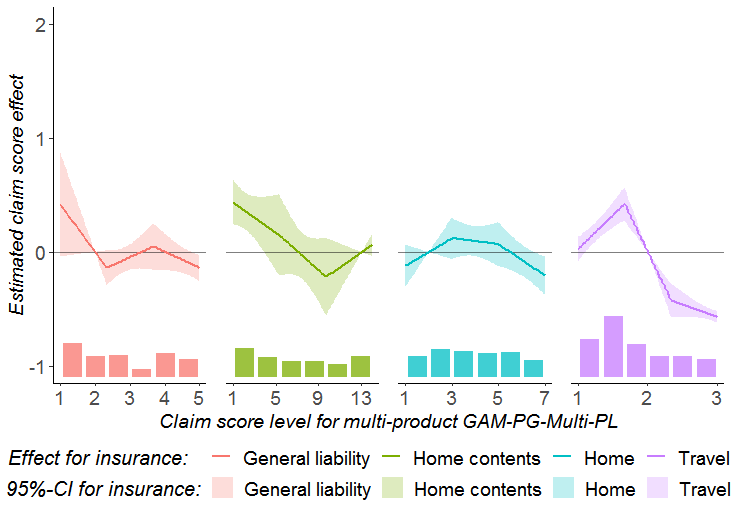}
            \caption{Effect on travel insurance}
            \label{F4.4.1d}
        \end{subfigure}
    \end{tabular}%\vspace{-5pt}
    \caption{Estimated claim score effects with corresponding $95\%$ confidence intervals and observed claim frequencies on general liability insurance (panel (a)), home contents insurance (panel (b)), home insurance (panel (c)) and travel insurance (panel (d)) for multi-product piecewise linear GAM-PG-Multi-PL.}
	\label{F4.4.1}
\end{figure}

While most claim scores in the multi-product GAMs lead to an intuitive and decreasing relation with respect to the risk of a customer, some are less straightforward and more complicated. However, in practice insurers must explain and justify their premia, and they thus highly prefer intuitive and interpretable premium rates. As a consequence, it makes more sense from a practical perspective to consider a rate structure segmented into piecewise linear components by using linear, rather than cubic, splines. We therefore implement this multi-product piecewise linear GAM for both Poisson and NB distributed claim frequencies and both Gamma and IG distributed claim severities, which we again abbreviate in \Cref{T4.4.1}. Moreover, we present the piecewise linear effects resulting from these claim scores in \Cref{F4.4.1} and show the maximal ratio Gini coefficients when including these piecewise linear specifications in \Cref{T4.4.2}. Finally, all parameter estimates are reported in \ref{AppendixC}.

\begin{table}[b!]%[ht!]
    \caption{Model abbreviations for different combinations of dynamic multivariate frequency and severity models, extended to piecewise linear specifications.}
    \label{T4.4.1}\vspace{-3pt}
	\centerline{\scalebox{0.80}{\begin{tabular}{l l l }
	    \toprule \addlinespace[1ex] \vspace{1pt}
		\textbf{Abbreviation} & \textbf{Frequency model} & \textbf{Severity model} \\ \hline \addlinespace[0.4ex]
		\textit{GAM-PG-Multi-PL} & Poisson multi-product claim score piecewise linear GAM & Gamma GLM \\
		\textit{GAM-PIG-Multi-PL} & Poisson multi-product claim score piecewise linear GAM & Inverse-Gaussian GLM \\
		\textit{GAM-NBG-Multi-PL} & Negative Binomial multi-product claim score piecewise linear GAM & Gamma GLM \\
		\textit{GAM-NBIG-Multi-PL} & Negative Binomial multi-product claim score piecewise linear GAM & Inverse-Gaussian GLM \\
		\bottomrule
	\end{tabular}}} %\vspace{10pt}
\end{table} %\newpage

%%% Multi-product splines %%%
 %\newpage%\clearpage

From the claim score effects in \Cref{F4.4.1}, we observe approximately the same patterns and subtleties for the piecewise linear GAMs as those for the cubic GAMs. In general, we again expect customers with lower claim scores on a certain product to be associated with more risk on that same product and that customers who merely possess home contents and/or travel insurance are associated with more risk for all other product categories. In terms of cross-selling opportunities, this also means that insurers should, for instance, not target customers holding home contents and/or travel insurance with cross-selling offers. Note that we only show the claim score effects for piecewise linear GAM-PG-Multi-PL in \Cref{F4.4.1} since the other three specifications lead to almost the same relations and that these effects are based on the optimal claim score parameters for the cubic GAMs. The piecewise linear specifications can, of course, be optimized separately as well, but this leads to similar results. The resulting piecewise linear splines therefore simply seem a piecewise linear simplification of the cubic splines and the subtleties in the claim scores indeed a data-driven result that the linear claim scores in the multi-product GLMs are unable to capture. As such, the piecewise linear GAMs represent a more intuitive and interpretable version of the cubic GAM for insurers to adopt in practice, but retain the possibility to identify the cross-selling potential of customers across property and casualty insurance.

Despite the promising potential of the piecewise linear GAMs, they do not significantly outperform their cubic counterparts in terms of our mini-max strategy. More specifically, in \Cref{T4.4.2} we find that the multi-product piecewise linear GAM performs comparable to the multi-product cubic GAM and typically outperforms the multi-product GLM, but that all other results remain effectively unaffected by these additional specifications. However, these piecewise linear GAMs mostly still seem, on average, much less vulnerable to alternative model choices than the static GLMs. In addition, we already found in the previous section that the multi-product model significantly outperforms the one-product model in all cubic GAM and some GLM cases in-sample based on a straightforward LR test. Even though these piecewise linear GAMs are not optimal in terms of our mini-max strategy, they therefore do seem promising to consider for practitioners in the non-life insurance industry.

%%% Ratio Gini coefficients and SE's %%%
\begin{table}[t!]%[ht!]
    \caption{Maximal ratio Gini coefficients in percentages with corresponding standard errors in parenthesis and rank for each product category with dynamic multivariate risk classification, extended to piecewise linear specifications.}
    \label{T4.4.2}\vspace{-3pt}
	\centerline{\scalebox{0.80}{\begin{tabular}{l r@{ }r@{}l r c r@{ }r@{}l r c r@{ }r@{}l r c r@{ }r@{}l r }
	    \toprule \addlinespace[1ex] \vspace{1pt}
	    & \multicolumn{4}{c}{\textbf{General liability}} &  & \multicolumn{4}{c}{\textbf{Home contents}} &  & \multicolumn{4}{c}{\textbf{Home}} &  & \multicolumn{4}{c}{\textbf{Travel}} \\
		\cline{2-5} \cline{7-10} \cline{12-15} \cline{17-20} \addlinespace[0.4ex]
		\textbf{Benchmark} & \multicolumn{3}{c}{\textbf{Coefficient}} & \textbf{Rank} &  & \multicolumn{3}{c}{\textbf{Coefficient}} & \textbf{Rank} &  & \multicolumn{3}{c}{\textbf{Coefficient}} & \textbf{Rank} &  & \multicolumn{3}{c}{\textbf{Coefficient}} & \textbf{Rank} \\ \hline \addlinespace[0.4ex]
		\textit{GLM-PG} & 15.96& (7.98)& & 21 &  & 18.16& (4.99)& & 11 &  & 12.38& (7.68)& & 10 &  & 33.12& (4.05)& & 23 \\
		\textit{GLM-PIG} & 15.96& (7.98)& & 20 &  & 24.13& (4.67)& & 24 &  & 10.42& (7.63)& & 8 &  & 32.42& (3.99)& & 21 \\
		\textit{GLM-NBG} & 15.96& (7.98)& & 19 &  & 18.09& (4.99)& & 10 &  & 12.32& (7.67)& & 9 &  & 33.14& (4.04)& & 24 \\
		\textit{GLM-NBIG} & 15.96& (7.98)& & 18 &  & 23.96& (4.67)& & 23 &  & 10.41& (7.63)& & 7 &  & 32.45& (3.98)& & 22 \\ \hline \addlinespace[0.2ex]
		\textit{GAM-PG-One} & 12.64& (7.36)& & 10 &  & 19.93& (4.47)& & 13 &  & 19.24& (7.78)& & 24 &  & 19.13& (3.18)& & 3 \\
		\textit{GAM-PIG-One} & 12.64& (7.36)& & 9 &  & 21.32& (4.44)& & 18 &  & 14.42& (7.62)& & 16 &  & \textbf{18.43}& \textbf{(3.42)}& & \textbf{1} \\
		\textit{GAM-NBG-One} & 12.78& (7.20)& & 12 &  & \textbf{7.79}& \textbf{(4.64)}& & \textbf{1} &  & 19.16& (7.77)& & 23 &  & 19.28& (3.20)& & 4 \\
		\textit{GAM-NBIG-One} & 12.78& (7.20)& & 11 &  & 19.37& (4.43)& & 12 &  & 14.40& (7.62)& & 15 &  & 18.54& (3.49)& & 2 \\
		\textit{GLM-PG-One} & 16.47& (8.04)& & 22 &  & 16.14& (4.92)& & 9 &  & 14.70& (7.76)& & 17 &  & 25.26& (4.14)& & 19 \\
		\textit{GLM-PIG-One} & 16.47& (8.04)& & 23 &  & 22.37& (4.68)& & 20 &  & 8.45& (7.65)& & 4 &  & 25.23& (4.14)& & 18 \\
		\textit{GLM-NBG-One} & 16.60& (8.03)& & 24 &  & 16.07& (4.91)& & 8 &  & 14.74& (7.76)& & 18 &  & 25.29& (4.20)& & 20 \\
		\textit{GLM-NBIG-One} & 15.94& (8.07)& & 17 &  & 22.12& (4.68)& & 19 &  & 8.43& (7.65)& & 3 &  & 25.20& (4.14)& & 17 \\ \hline \addlinespace[0.2ex]
		\textit{GAM-PG-Multi} & 6.56& (8.04)& & 4 &  & 10.35& (4.72)& & 4 &  & 15.28& (7.67)& & 19 &  & 19.97& (3.24)& & 9 \\
		\textit{GAM-PIG-Multi} & 6.56& (8.04)& & 3 &  & 23.49& (4.40)& & 22 &  & 12.68& (7.58)& & 12 &  & 19.47& (3.39)& & 5 \\
		\textit{GAM-NBG-Multi} & \textbf{6.56}& \textbf{(8.04)}& & \textbf{1} &  & 8.45& (4.80)& & 3 &  & 15.28& (7.67)& & 20 &  & 20.11& (3.26)& & 11 \\
		\textit{GAM-NBIG-Multi} & 6.56& (8.04)& & 2 &  & 23.19& (4.40)& & 21 &  & 12.67& (7.58)& & 11 &  & 19.58& (3.41)& & 6 \\
		\textit{GLM-PG-Multi} & 14.44& (7.98)& & 13 &  & 16.04& (4.93)& & 7 &  & 9.88& (7.64)& & 6 &  & 24.78& (4.20)& & 15 \\
		\textit{GLM-PIG-Multi} & 14.44& (7.98)& & 14 &  & 20.90& (4.82)& & 17 &  & 7.12& (7.61)& & 2 &  & 24.17& (4.17)& & 14 \\
		\textit{GLM-NBG-Multi} & 14.47& (7.97)& & 15 &  & 15.94& (4.92)& & 6 &  & 9.86& (7.64)& & 5 &  & 24.78& (4.20)& & 16 \\
		\textit{GLM-NBIG-Multi} & 14.94& (8.04)& & 16 &  & 20.84& (4.81)& & 16 &  & \textbf{6.94}& \textbf{(7.59)}& & \textbf{1} &  & 24.16& (4.15)& & 13 \\ \hline \addlinespace[0.2ex]
		\textit{GAM-PG-Multi-PL} & 6.75& (8.05)& & 6 &  & 11.61& (4.77)& & 5 &  & 15.42& (7.68)& & 22 &  & 20.08& (3.23)& & 10 \\
		\textit{GAM-PIG-Multi-PL} & 6.75& (8.05)& & 5 &  & 20.64& (4.40)& & 15 &  & 13.33& (7.61)& & 14 &  & 19.59& (3.38)& & 7 \\
		\textit{GAM-NBG-Multi-PL} & 6.77& (8.05)& & 8 &  & 8.39& (4.80)& & 2 &  & 15.41& (7.68)& & 21 &  & 20.21& (3.24)& & 12 \\
		\textit{GAM-NBIG-Multi-PL} & 6.77& (8.05)& & 7 &  & 19.93& (4.41)& & 14 &  & 13.31& (7.61)& & 13 &  & 19.70& (3.40)& & 8 \\
		\bottomrule
	\end{tabular}}}
\end{table} %\newpage

The claims experience of customers thus appears to be an important determinant for individual risk classification and it can be very profitable to account for this experience in our premia. Moreover, it seems that accounting for multi-product claims experience is only optimal in case of a portfolio with relatively many policyholders having claims on multiple products and with sufficient overlap between different product categories. In all other cases, it suffices to only incorporate the claims experience in the own product category. However, if our insurance portfolio does satisfy these conditions, the multi-product piecewise linear GAM seems particularly interesting for its intuitive and interpretable use in practice, and its ability to detect the cross-selling potential of existing individual customers. % 4 Applications in non-life insurance
\section{Conclusion} \label{Section5}

In this paper, we have presented and applied a multi-product framework for dynamic insurance pricing on the level of individual policyholders. While the industry standard of a GLM typically considers only \textit{a priori} information of policyholders, we have included the \textit{a posteriori} claims experience of customers across multiple product categories in a predictive claim score. As such, we have extended the BMS-panel model of \citet{boucher2014} and \citet{boucher2018} by on the one hand incorporating the claims experience from multiple product lines and on the other hand allowing the respective claim scores to have a non-linear effect on the (logarithm of the) premium rate structure. Moreover, we have considered both a natural cubic and linear spline for the effects of these claim scores to embed our novel multi-product framework into a GAM and benefit from its existing framework.

In our application of this multi-product framework, we considered a Dutch property and casualty insurance portfolio, including general liability, home contents, home and travel insurance. Using this portfolio, we compared the industry standard of a GLM with non-linear cubic splines for the claim scores and linear effects similar to the BMS-panel model. This led to the finding that accounting for a customer's claims experience can be very profitable and substantially outperforms a static GLM based on a mini-max strategy of ratio Gini coefficients. This mini-max strategy generally also favored the cubic splines more than the linear effects in terms of profit potential, where the linear effects appeared to be dominated by the good risks and thus misspecified for the other risks. The effects from the cubic splines, on the other hand, were actually a data-driven result and yielded more representative confidence bounds in case of claim score levels with little exposure. A piecewise linear simplification of the cubic spline supported this claim and resulted in almost the same claim score effects and identified subtle cross-selling opportunities that the linear specification was unable to detect. More importantly, however, our results seemed to indicate that, in case of a portfolio with relatively many policyholders having claims on multiple products and with sufficient overlap between different product lines, it is in fact optimal or most profitable to account for the claims experience of a customer from all product categories. As such, the multi-product framework presented in this paper, and in particular the piecewise linear GAMs for their intuitive and interpretable rate structures, seem promising for practitioners in the non-life insurance industry to implement in their dynamic pricing strategies.

While the focus of this paper has primarily been on separate effects for each claim score, it is also possible to include interaction effects of all these scores. However, a more interesting avenue for future research is to consider a single multi-dimensional spline in the multi-product GAM for all the separate claim scores combined. This, in turn, may be able to expose complex dependencies between the claim scores of different product categories and enhance the multi-product risk profiles. Alternatively, future research can refine the piecewise linear simplification of the cubic spline by using one of the binning strategies mentioned in \citet{henckaerts2018} or by adopting a monotonicity restriction on the spline. Both refinements may improve the profitability of the multi-product piecewise linear GAM and may lead to a more intuitively appealing framework for non-life insurers to adopt in practice. % Conclusion

%\newpage
\section*{Acknowledgements}

The author gratefully acknowledges financial support from VIVAT insurance. Any errors made or views expressed in this paper are the responsibility of the author alone. % Acknowledgements

\begingroup
    \setlength{\bibsep}{
    \if1\double
        7.5pt
    \else
        0.0pt
    \fi}
    \def\bibfont{\small}
    \bibliography{References_Paper} % References
\endgroup

%\newpage
\appendix
{\renewcommand{\thesection}{Appendix \Alph{section}}
\renewcommand{\theequation}{\Alph{section}.\arabic{equation}}
\section{Risk factors for property and casualty insurance} \label{AppendixA}

\begin{table}[ht!]
    \caption{Description of the key variables used for property and casualty insurance.}
    \label{TA.1}\vspace{-3pt}
	\centerline{\scalebox{1.00}{\begin{tabularx}{\textwidth}{l l l }
		\toprule \addlinespace[1ex] \vspace{1pt}
		\textbf{Variable}\hspace{46pt} & \textbf{Values}\hspace{25pt} & \textbf{Description} \\ \hline \addlinespace[0.4ex]
		\texttt{Count} & Integer & The number of claims filed by the policyholder. \\
		\texttt{Size} & Continuous & The size of the claim in euro's.  \\
		\texttt{Exposure} & Continuous & The exposure to risk in years. \\
		\bottomrule
	\end{tabularx}}}
\end{table} %\newpage

\begin{table}[ht!]
    \caption{Description of the risk factors used for general liability insurance.}
    \label{TA.2}\vspace{-3pt}
	\centerline{\scalebox{1.00}{\begin{tabularx}{\textwidth}{l l l }
		\toprule \addlinespace[1ex] \vspace{1pt}
		\textbf{Risk factor}\hspace{31pt} & \textbf{Values}\hspace{25pt} & \textbf{Description} \\ \hline \addlinespace[0.4ex]
		\texttt{FamilySituation} & $4$ categories & Type of family situation. \\
		\bottomrule
	\end{tabularx}}}
\end{table} %\newpage

\begin{table}[ht!]
    \caption{Description of the risk factors used for home contents insurance.}
    \label{TA.3}\vspace{-3pt}
	\centerline{\scalebox{1.00}{\begin{tabularx}{\textwidth}{l l l }
		\toprule \addlinespace[1ex] \vspace{1pt}
		\textbf{Risk factor}\hspace{31pt} & \textbf{Values} & \textbf{Description} \\ \hline \addlinespace[0.4ex]
		\texttt{ProductType} & $3$ categories & Type of product. \\
		\texttt{Year} & $6$ categories & Calendar year. \\
		\texttt{Age} & Continuous & Age of the policyholder in years. \\
		\texttt{BuildingType} & $10$ categories & Type of building of the home. \\
		\texttt{RoofType} & $5$ categories & Type of roof of the home. \\
		\texttt{FloorSpace} & Continuous & Total floor area of the home in thousands of square metres.  \\
		\texttt{HomeOwner} & $4$ categories & Whether the policyholder owns or rents its home. \\
		\texttt{Residence} & $5$ categories & Residential area of the policyholder. \\
		\texttt{Urban} & $8$ categories & Degree of urbanisation at home address. \\
		\texttt{GlassCoverage} & $2$ categories & Whether the policyholder has glass coverage. \\
		\bottomrule
	\end{tabularx}}}
\end{table} %\newpage

\begin{table}[ht!]
    \caption{Description of the risk factors used for home insurance.}
    \label{TA.4}\vspace{-3pt}
	\centerline{\scalebox{1.00}{\begin{tabularx}{\textwidth}{l l l }
		\toprule \addlinespace[1ex] \vspace{1pt}
		\textbf{Risk factor}\hspace{26pt} & \textbf{Values} & \textbf{Description} \\ \hline \addlinespace[0.4ex]
		\texttt{ProductType} & $3$ categories & Type of product. \\
		\texttt{Year} & $6$ categories & Calendar year. \\
		\texttt{Age} & Continuous & Age of the policyholder in years. \\
		\texttt{FamilySituation} & $5$ categories & Type of family situation. \\
		\texttt{BuildingType} & $10$ categories & Type of building of the home. \\
		\texttt{RoofType} & $3$ categories & Type of roof of the home. \\
		\texttt{Capacity} & Continuous & Total capacity of the home in thousands of cubic metres.  \\
		\texttt{ConstructionYear} & $6$ categories & Construction year of the home. \\
		\texttt{Residence} & $5$ categories & Residential area of the policyholder. \\
		\texttt{Urban} & $8$ categories & Degree of urbanisation at home address. \\
		\texttt{GlassCoverage} & $2$ categories & Whether the policyholder has glass coverage. \\
		\bottomrule
	\end{tabularx}}}
\end{table} \newpage

\begin{table}[ht!]
    \caption{Description of the risk factors used for travel insurance.}
    \label{TA.5}\vspace{-3pt}
	\centerline{\scalebox{1.00}{\begin{tabularx}{\textwidth}{l l l }
		\toprule \addlinespace[1ex] \vspace{1pt}
		\textbf{Risk factor} & \textbf{Values}\hspace{25pt} & \textbf{Description} \\ \hline \addlinespace[0.4ex]
		\texttt{Region} & $3$ categories & Regional area covered. \\
		\texttt{Age} & Continuous & Age of the policyholder in years. \\
		\texttt{FamilySituation} & $5$ categories & Type of family situation. \\
		\texttt{WinterCoverage} & $2$ categories & Whether the policyholder has winter sport coverage. \\
		\texttt{MoneyCoverage} & $2$ categories & Whether the policyholder has money coverage. \\
		\texttt{VehicleCoverage} & $2$ categories & Whether the policyholder has vehicle coverage. \\
		\texttt{MedicalCoverage} & $2$ categories & Whether the policyholder has medical coverage. \\
		\texttt{AccidentCoverage} & $2$ categories & Whether the policyholder has accident coverage. \\
		\texttt{CancelCoverage} & $2$ categories & Whether the policyholder has cancellation coverage. \\
		\bottomrule
	\end{tabularx}}}
\end{table} %\newpage % Appendix A Risk factors for property and casualty insurance
\section{Estimation in multi-product claim score model} \label{AppendixB}

Essential to the multi-product claim score model developed in this paper is the assumption that the response variable is independently distributed according to some member of the exponential family. If we denote by $Y_{i,t}$ this response for subject $i$ in period $t$, then its density $p(\cdot)$ can be written as
\begin{equation} \label{EB.1}
    p(y_{i,t}|\vartheta_{i,t}, \varphi) = h(y_{i,t}, w_{i,t}, \varphi) \exp \left( \frac{w_{i,t}}{\varphi} \left( \vartheta_{i,t} y_{i,t} - A(\vartheta_{i,t}) \right) \right) \quad \text{for} \quad i = 1, \dots, M, \ t = 1, \dots, T_i,
\end{equation}
where $\vartheta_{i,t}$ denotes a distribution parameter, $\varphi$ a dispersion parameter, $w_{i,t}$ a known weight that is typically set to one or the exposure to risk, $h(\cdot, \cdot, \cdot)$ a known function and $A(\cdot)$ a known twice continuously differentiable function. While the function $h(\cdot, \cdot, \cdot)$ is of little interest in GLM theory, the function $A(\cdot)$ is related to the mean $\mu_{i,t}$ and covariance $\Sigma_{i,t}$ of the response variable through
\begin{equation*}
    \mu = \mathbb{E}\left[ Y \right] = A'(\vartheta) \quad \text{and} \quad \Sigma = \mathbb{V}\left[ Y \right] = \frac{\varphi}{w} v(\mu),
\end{equation*}
where $v(\mu) = A''(\vartheta) = A''(A'^{-1}(\mu))$ is called the variance function \citep{ohlsson2010}. As such, it is sufficient to only consider a model for the mean since this can already completely characterize the entire distribution of the response variable.

When considering the multi-product claim score model for the mean equation, inference can be performed straightforwardly by penalized Maximum Likelihood. Suppose the linear predictor in this model is given by \Cref{E2.4} with penalized cubic regression splines for the transformations $f^{(c)}_{j}(\cdot)$. If we consider $(m + 1)$-th order B-splines with $k$ parameters and $k + m + 1$ knots in a certain interval $[ x_1, x_{k + m + 1}]$ for the basis of some set of regression splines, then we can represent them by
\begin{equation*}
    f_{j}(x) = \sum_{h = 1}^{k} \gamma_{h} B^{m}_{h}(x) \quad \text{for} \quad j = 1, \dots, C,
\end{equation*}
with $m = 2$ for cubic splines and $m = 0$ for linear splines, and where the B-spline basis functions $B^{m}_{h}(\cdot)$ are defined recursively as
\begin{equation*}
    B^{m}_{h}(x) = \frac{x - x_{h}}{x_{h + m + 1} - x_{h}} B^{m - 1}_{h}(x) + \frac{x_{h + m + 2} - x}{x_{h + m + 2} - x_{h + 1}} B^{m - 1}_{h + 1}(x) \quad \text{for} \quad h = 1, \dots, k
\end{equation*}
with initial value $B^{-1}_{h}(x) = \mathbbm{1}\left(x_{h} \leq x < x_{h+1}\right)$ and where we have omitted the superscripts $(c)$ for the sake of simplicity \citep{wood2006}. These smooths $f_{j}(\cdot)$ are usually subject to an additional centering constraint to ensure identification of the mean equation and typically it is assumed that all its elements sum to zero. As a result, one degree of freedom in the splines is lost due to this identification restriction and $k-1$ effectively remain. The penalized log-likelihood function is now defined as
\begin{align}\label{EB.2}\nonumber
    \ell_{p}(\delta, \varphi, \lambda|y) &= \ell(\delta, \varphi|y) - \frac{1}{2} \sum_{j = 1}^{C} \lambda_{j} \int f_{j}^{\prime\prime}(x)^2 \mathrm{d}x \\
    &= \sum_{i = 1}^{M} \sum_{t = 1}^{T_i} \left[ \log\left( h(y_{i,t}, w_{i,t}, \varphi) \right) + \frac{w_{i, t}}{\varphi} \left(\vartheta_{i,t} y_{i,t} - A(\vartheta_{i,t})\right) \right] - \frac{1}{2} \sum_{j = 1}^{C} \lambda_{j} \delta^{\prime} S_{j} \delta,
\end{align}
with $\delta = (\beta, \gamma)$ and where the distribution parameters $\vartheta_{i,t}$ depend on the parameters $\delta$ through the linear predictor, $\lambda_{j}$ denotes the penalty or smoothing parameter for the $j$-th regression spline and $S_{j}$ a matrix of known coefficients $\tilde{S}_{j}$ padded with zeros such that $\delta^{\prime} S_{j} \delta = \gamma^{\prime} \tilde{S}_{j} \gamma$. Note that the first expression in \Cref{EB.2}, or $\ell(\cdot)$, actually represents the ordinary log-likelihood function of the model and that the multi-product claim score model can therefore be seen as a penalized GLM in terms of optimization. Maximization of this penalized log-likelihood in terms of the parameters $\delta$ given the penalties $\lambda_{j}$ leads to the set of $K + \sum_{j = 1}^{C} k_{j}$ normal equations given by
\begin{equation}\label{EB.3}
    \frac{1}{\varphi} \sum_{i = 1}^{M} \sum_{t = 1}^{T_i} w_{i,t} \frac{y_{i,t} - \mu_{i,t}}{v(\mu_{i,t}) g'(\mu_{i,t})} X_{i,t} - \sum_{j = 1}^{C} \lambda_{j} S_{j} \delta = 0,
\end{equation}
where $K$ denotes the dimension of $\beta$ and $\varphi$ is usually omitted since we can incorporate its effect into the penalties. In practice, the smoothing parameters are of course unknown as well and are usually estimated by generalized cross-validation or unbiased risk estimation (see, e.g., \citet{wood2006}). 

It is clear that these normal equations do not lead to an analytical solution for our unknown parameters and that we need to find a numerical solution to them. One way to numerically solve these equations is by the Newton-Raphson method that relies on the gradient of the normal equations with respect to $\delta$, or the Hessian matrix of the (penalized) log-likelihood function. However, a more popular approach for numerically solving these equations in the context of GLMs is called the Fisher scoring method. This method applies the same iterative procedure as the Newton-Raphson method, but now uses the Fisher information matrix $\mathcal{I}(\cdot)$, rather than the Hessian matrix. The Fisher scoring method is therefore characterized by
\begin{equation} \label{EB.4}
    \delta^{(n+1)} = \delta^{(n)} + \mathcal{I}^{-1}(\delta^{(n)}) J(\delta^{(n)}),
\end{equation}
with $J(\cdot)$ the Jacobian matrix of the (penalized) log-likelihood function, or the normal equations. Formally, this information matrix is given by the expectation of the negative Hessian matrix, or
\begin{equation}\label{EB.5}
    \mathcal{I}(\delta) = \mathbb{E}\left[ -H(\delta) \right] = \frac{1}{\varphi} \sum_{i = 1}^{M} \sum_{t = 1}^{T_i} \frac{w_{i,t}}{v(\mu_{i,t}) g'(\mu_{i,t})^2} X_{i,t} X_{i,t}^{\prime} - \sum_{j = 1}^{C} \lambda_{j} S_{j},
\end{equation}
where $\varphi$ is typically omitted again. The advantages of using this matrix are that it is slightly easier to implement in practice and, by definition, always remains positive definite \citep{ohlsson2010}. The Hessian matrix, on the other hand, is not necessarily positive definite unless we are already close to convergence. The Fisher scoring method therefore typically leads to more stable convergence than the Newton-Raphson method, whereas the latter method is considered faster. In the context of (penalized) GLMs, Fisher's iterative procedure is also known as (Penalized) Iteratively Re-weighted Least Squares and can easily be implemented in, for instance, \texttt{R} with the package \texttt{mgcv} developed by \citet{wood2006}. As such, the multi-product claim score model can heavily benefit from the framework of GLMs and GAMs, and rely on existing statistical theory and software for inference.  % Appendix B Estimation in multi-product claim score model
%\vspace{-10pt}
\section{Supplementary estimation results} \label{AppendixC}

%%%%%%%%%%%%%%%%%% Static univariate %%%%%%%%%%%%%%%%
%\vspace{-10pt}
\begin{table}[ht!]%[ht!]
    \caption{Parameter estimates and corresponding standard errors in parenthesis for general liability insurance with static univariate risk classification.}
    \label{T.C.1}\vspace{-3pt}
	\centerline{\scalebox{0.80}{
\begin{tabular}{l r@{ }r@{}l r@{ }r@{}l c r@{ }r@{}l r@{ }r@{}l r@{ }r@{}l }
	    \toprule \addlinespace[1ex] \vspace{1pt}
	    & \multicolumn{6}{c}{\textbf{Frequency GLM}} &  & \multicolumn{6}{c}{\textbf{Severity GLM}} \\
	    \cline{2-7} \cline{9-14} \addlinespace[0.4ex]
		\textbf{Risk factor} & \multicolumn{3}{c}{\textbf{Poisson}} & \multicolumn{3}{c}{\textbf{Negative Binomial}} &  & \multicolumn{3}{c}{\textbf{Gamma}} & \multicolumn{3}{c}{\textbf{Inverse-Gaussian}} \\ \hline \addlinespace[0.4ex]
		\textit{\texttt{Constant}} & -4.2961& (0.0408)&$^{***}$ & -4.2961& (0.0410)&$^{***}$ &  & 6.3688& (0.2493)&$^{***}$ & 6.3688& (0.2089)&$^{***}$ \\
		\textit{\texttt{FamilySituation}} & && & && &  & && & && \\
		\hspace{10pt}\textit{- Category 1} & 0.0960& (0.0533)& & 0.0960& (0.0536)& &  & 0.5036& (0.3253)& & 0.5036& (0.3072)& \\
		\hspace{10pt}\textit{- Category 2} & 1.0779& (0.0498)&$^{***}$ & 1.0780& (0.0502)&$^{***}$ &  & 0.3811& (0.3040)& & 0.3811& (0.2735)& \\
		\hspace{10pt}\textit{- Category 3} & 1.0513& (0.0689)&$^{***}$ & 1.0515& (0.0697)&$^{***}$ &  & 0.1904& (0.4209)& & 0.1904& (0.3760)& \\
		\hline \addlinespace[0.4ex]
		\textit{Dispersion} & \multicolumn{3}{c}{1.0000} & \multicolumn{3}{c}{0.7909} & & \multicolumn{3}{c}{37.2808} & \multicolumn{3}{c}{0.0449} \\
		\bottomrule
		\multicolumn{14}{l}{\footnotesize Significance levels: $^{*}5\%$-level, $^{**}1\%$-level, $^{***}0.1\%$-level or less} \\
	\end{tabular}}}
\end{table} %\newpage

\begin{table}[ht!]%[ht!]
    \caption{Parameter estimates and corresponding standard errors in parenthesis for home contents insurance with static univariate risk classification.}
    \label{T.C.2}\vspace{-3pt}
	\centerline{\scalebox{0.80}{\begin{tabular}{l r@{ }r@{}l r@{ }r@{}l c r@{ }r@{}l r@{ }r@{}l }
	    \toprule \addlinespace[1ex] \vspace{1pt}
	    & \multicolumn{6}{c}{\textbf{Frequency GLM}} &  & \multicolumn{6}{c}{\textbf{Severity GLM}} \\
	    \cline{2-7} \cline{9-14} \addlinespace[0.4ex]
		\textbf{Risk factor} & \multicolumn{3}{c}{\textbf{Poisson}} & \multicolumn{3}{c}{\textbf{Negative Binomial}} &  & \multicolumn{3}{c}{\textbf{Gamma}} & \multicolumn{3}{c}{\textbf{Inverse-Gaussian}} \\ \hline \addlinespace[0.4ex]
		\textit{\texttt{Constant}} & -2.3702& (0.0709)&$^{***}$ & -2.3670& (0.0721)&$^{***}$ &  & 6.7431& (0.1582)&$^{***}$ & 6.7820& (0.1578)&$^{***}$ \\
		\textit{\texttt{ProductType}} & && & && &  & && & && \\
		\hspace{10pt}\textit{- Category 1} & -0.5310& (0.1503)&$^{***}$ & -0.5325& (0.1520)&$^{***}$ &  & 0.2834& (0.3134)& & 0.2865& (0.3430)& \\
		\hspace{10pt}\textit{- Category 2} & -0.1088& (0.0467)& & -0.1091& (0.0474)& &  & 0.0316& (0.1022)& & 0.0410& (0.0990)& \\
		\textit{\texttt{Year}} & && & && &  & && & && \\
		\hspace{10pt}\textit{- Category 1} & -0.0019& (0.0662)& & -0.0030& (0.0672)& &  & 0.0482& (0.1434)& & 0.0145& (0.1401)& \\
\end{tabular}}}
\end{table} %\newpage

\begin{table}[ht!]%[ht!]
    \caption{Parameter estimates and corresponding standard errors in parenthesis for home insurance with static univariate risk classification.}
    \label{T.C.3}\vspace{-3pt}
	\centerline{\scalebox{0.80}{\begin{tabular}{l r@{ }r@{}l r@{ }r@{}l c r@{ }r@{}l r@{ }r@{}l }
	    \toprule \addlinespace[1ex] \vspace{1pt}
	    & \multicolumn{6}{c}{\textbf{Frequency GLM}} &  & \multicolumn{6}{c}{\textbf{Severity GLM}} \\
	    \cline{2-7} \cline{9-14} \addlinespace[0.4ex]
		\textbf{Risk factor} & \multicolumn{3}{c}{\textbf{Poisson}} & \multicolumn{3}{c}{\textbf{Negative Binomial}} &  & \multicolumn{3}{c}{\textbf{Gamma}} & \multicolumn{3}{c}{\textbf{Inverse-Gaussian}} \\ \hline \addlinespace[0.4ex]
		\textit{\texttt{Constant}} & -2.1048& (0.1125)&$^{***}$ & -2.1052& (0.1136)&$^{***}$ &  & 7.5419& (0.2097)&$^{***}$ & 7.5565& (0.2166)&$^{***}$ \\
		\textit{\texttt{ProductType}} & && & && &  & && & && \\
		\hspace{10pt}\textit{- Category 1} & -0.4486& (0.2018)&$^{*}$ & -0.4451& (0.2033)&$^{*}$ &  & -0.1019& (0.3817)& & -0.0798& (0.3680)& \\
		\hspace{10pt}\textit{- Category 2} & -0.3316& (0.0681)&$^{***}$ & -0.3308& (0.0686)&$^{***}$ &  & -0.2174& (0.1343)& & -0.2324& (0.1496)& \\
		\textit{\texttt{Year}} & && & && &  & && & && \\
		\hspace{10pt}\textit{- Category 1} & -0.3053& (0.0980)&$^{**}$ & -0.3046& (0.0989)&$^{**}$ &  & -0.3758& (0.1890)&$^{*}$ & -0.3885& (0.1993)& \\
\end{tabular}}}
\end{table} %\newpage

\begin{table}[ht!]%[ht!]
    \caption{Parameter estimates and corresponding standard errors in parenthesis for travel insurance with static univariate risk classification.}
    \label{T.C.4}\vspace{-3pt}
	\centerline{\scalebox{0.80}{\begin{tabular}{l r@{ }r@{}l r@{ }r@{}l c r@{ }r@{}l r@{ }r@{}l }
	    \toprule \addlinespace[1ex] \vspace{1pt}
	    & \multicolumn{6}{c}{\textbf{Frequency GLM}} &  & \multicolumn{6}{c}{\textbf{Severity GLM}} \\
	    \cline{2-7} \cline{9-14} \addlinespace[0.4ex]
		\textbf{Risk factor} & \multicolumn{3}{c}{\textbf{Poisson}} & \multicolumn{3}{c}{\textbf{Negative Binomial}} &  & \multicolumn{3}{c}{\textbf{Gamma}} & \multicolumn{3}{c}{\textbf{Inverse-Gaussian}} \\ \hline \addlinespace[0.4ex]
		\textit{\texttt{Constant}} & -3.2927& (0.0480)&$^{***}$ & -3.2906& (0.0485)&$^{***}$ &  & 5.5791& (0.1295)&$^{***}$ & 5.5862& (0.1258)&$^{***}$ \\
		\textit{\texttt{Region}} & && & && &  & && & && \\
		\hspace{10pt}\textit{- Category 1} & 0.3307& (0.0293)&$^{***}$ & 0.3311& (0.0297)&$^{***}$ &  & -0.1336& (0.0749)& & -0.1410& (0.0712)&$^{*}$ \\
		\hspace{10pt}\textit{- Category 2} & 0.7045& (0.0264)&$^{***}$ & 0.7067& (0.0268)&$^{***}$ &  & 0.0493& (0.0764)& & 0.0497& (0.0766)& \\
		\textit{\texttt{Age}} & -0.0052& (0.0008)&$^{***}$ & -0.0052& (0.0008)&$^{***}$ &  & 0.0049& (0.0023)&$^{*}$ & 0.0047& (0.0023)&$^{*}$ \\
		\textit{\texttt{FamilySituation}} & && & && &  & && & && \\
\end{tabular}}}
\end{table} %\newpage

%%%%%%%%%%%%%%%%%% Dynamic univariate %%%%%%%%%%%%%%%%
\begin{table}[ht!]%[ht!]
    \caption{Parameter estimates and corresponding standard errors in parenthesis for general liability insurance with dynamic univariate risk classification.}
    \label{T.C.6}\vspace{-3pt}
	\centerline{\scalebox{0.525}{\begin{tabular}{l r@{ }r@{}l r@{ }r@{}l r@{ }r@{}l r@{ }r@{}l r@{ }r@{}l r@{ }r@{}l r@{ }r@{}l r@{ }r@{}l }
	    \toprule \addlinespace[1ex] \vspace{1pt}
	    & \multicolumn{24}{c}{\textbf{Claim frequency}} \\
	    \cline{2-25} \addlinespace[0.4ex]
		\textbf{Risk factor} & \multicolumn{3}{c}{\textbf{GAM-PG-One}} & \multicolumn{3}{c}{\textbf{GAM-PIG-One}} & \multicolumn{3}{c}{\textbf{GAM-NBG-One}} & \multicolumn{3}{c}{\textbf{GAM-NBIG-One}} & \multicolumn{3}{c}{\textbf{GLM-PG-One}} & \multicolumn{3}{c}{\textbf{GLM-PIG-One}} & \multicolumn{3}{c}{\textbf{GLM-NBG-One}} & \multicolumn{3}{c}{\textbf{GLM-NBIG-One}} \\ \hline \addlinespace[0.4ex]
		\textit{\texttt{Constant}} & -3.9454& (0.0567)&$^{***}$ & -3.9454& (0.0567)&$^{***}$ & -3.9458& (0.0572)&$^{***}$ & -3.9458& (0.0572)&$^{***}$ & -4.1020& (0.0424)&$^{***}$ & -4.1020& (0.0424)&$^{***}$ & -4.0997& (0.0427)&$^{***}$ & -4.0524& (0.0436)&$^{***}$ \\
		\textit{\texttt{FamilySituation}} & && & && & && & && & && & && & && & && \\
		\hspace{10pt}\textit{- Category 1} & 0.0718& (0.0533)& & 0.0718& (0.0533)& & 0.0721& (0.0536)& & 0.0721& (0.0536)& & 0.0847& (0.0533)& & 0.0847& (0.0533)& & 0.0848& (0.0535)& & 0.0829& (0.0535)& \\
		\hspace{10pt}\textit{- Category 2} & 1.0440& (0.0499)&$^{***}$ & 1.0440& (0.0499)&$^{***}$ & 1.0446& (0.0503)&$^{***}$ & 1.0446& (0.0503)&$^{***}$ & 1.0078& (0.0502)&$^{***}$ & 1.0078& (0.0502)&$^{***}$ & 1.0079& (0.0506)&$^{***}$ & 1.0104& (0.0506)&$^{***}$ \\
		\hspace{10pt}\textit{- Category 3} & 0.9677& (0.0693)&$^{***}$ & 0.9677& (0.0693)&$^{***}$ & 0.9672& (0.0701)&$^{***}$ & 0.9672& (0.0701)&$^{***}$ & 0.9606& (0.0694)&$^{***}$ & 0.9606& (0.0694)&$^{***}$ & 0.9609& (0.0701)&$^{***}$ & 0.9566& (0.0701)&$^{***}$ \\
		\hline \addlinespace[0.4ex]
		$\mathit{\gamma_{1}}$ & -0.4778& (0.0993)&$^{***}$ & -0.4778& (0.0993)&$^{***}$ & -0.4770& (0.1013)&$^{***}$ & -0.4770& (0.1013)&$^{***}$ & -0.0939& (0.0061)&$^{***}$ & -0.0939& (0.0061)&$^{***}$ & -0.0953& (0.0062)&$^{***}$ & -0.1147& (0.0075)&$^{***}$ \\
\end{tabular}}}
\end{table} %\newpage

\begin{table}[ht!]%[ht!]
    \caption{Parameter estimates and corresponding standard errors in parenthesis for home contents insurance with dynamic univariate risk classification.}
    \label{T.C.7}\vspace{-3pt}
	\centerline{\scalebox{0.465}{\begin{tabular}{l r@{ }r@{}l r@{ }r@{}l r@{ }r@{}l r@{ }r@{}l r@{ }r@{}l r@{ }r@{}l r@{ }r@{}l r@{ }r@{}l }
	    \toprule \addlinespace[1ex] \vspace{1pt}
	    & \multicolumn{24}{c}{\textbf{Claim frequency}} \\
	    \cline{2-25} \addlinespace[0.4ex]
		\textbf{Risk factor} & \multicolumn{3}{c}{\textbf{GAM-PG-One}} & \multicolumn{3}{c}{\textbf{GAM-PIG-One}} & \multicolumn{3}{c}{\textbf{GAM-NBG-One}} & \multicolumn{3}{c}{\textbf{GAM-NBIG-One}} & \multicolumn{3}{c}{\textbf{GLM-PG-One}} & \multicolumn{3}{c}{\textbf{GLM-PIG-One}} & \multicolumn{3}{c}{\textbf{GLM-NBG-One}} & \multicolumn{3}{c}{\textbf{GLM-NBIG-One}} \\ \hline \addlinespace[0.4ex]
		\textit{\texttt{Constant}} & -2.3967& (0.0729)&$^{***}$ & -2.3967& (0.0729)&$^{***}$ & -2.3943& (0.0741)&$^{***}$ & -2.3943& (0.0741)&$^{***}$ & -2.2890& (0.0712)&$^{***}$ & -2.2890& (0.0712)&$^{***}$ & -2.2844& (0.0723)&$^{***}$ & -2.2844& (0.0723)&$^{***}$ \\
		\textit{\texttt{ProductType}} & && & && & && & && & && & && & && & && \\
		\hspace{10pt}\textit{- Category 1} & -0.4917& (0.1507)&$^{**}$ & -0.4917& (0.1507)&$^{**}$ & -0.4929& (0.1522)&$^{**}$ & -0.4929& (0.1522)&$^{**}$ & -0.5765& (0.1502)&$^{***}$ & -0.5765& (0.1502)&$^{***}$ & -0.5781& (0.1518)&$^{***}$ & -0.5781& (0.1518)&$^{***}$ \\
		\hspace{10pt}\textit{- Category 2} & -0.1026& (0.0468)&$^{*}$ & -0.1026& (0.0468)&$^{*}$ & -0.1029& (0.0475)&$^{*}$ & -0.1029& (0.0475)&$^{*}$ & -0.1424& (0.0469)&$^{**}$ & -0.1424& (0.0469)&$^{**}$ & -0.1415& (0.0476)&$^{**}$ & -0.1415& (0.0476)&$^{**}$ \\
		\textit{\texttt{Year}} & && & && & && & && & && & && & && & && \\
		\hspace{10pt}\textit{- Category 1} & 0.0343& (0.0663)& & 0.0343& (0.0663)& & 0.0338& (0.0674)& & 0.0338& (0.0674)& & -0.0053& (0.0663)& & -0.0053& (0.0663)& & -0.0063& (0.0673)& & -0.0063& (0.0673)& \\
\end{tabular}}}
\end{table} %\newpage

\begin{table}[ht!]%[ht!]
    \caption{Parameter estimates and corresponding standard errors in parenthesis for home insurance with dynamic univariate risk classification.}
    \label{T.C.8}\vspace{-3pt}
	\centerline{\scalebox{0.460}{\begin{tabular}{l r@{ }r@{}l r@{ }r@{}l r@{ }r@{}l r@{ }r@{}l r@{ }r@{}l r@{ }r@{}l r@{ }r@{}l r@{ }r@{}l }
	    \toprule \addlinespace[1ex] \vspace{1pt}
	    & \multicolumn{24}{c}{\textbf{Claim frequency}} \\
	    \cline{2-25} \addlinespace[0.4ex]
		\textbf{Risk factor} & \multicolumn{3}{c}{\textbf{GAM-PG-One}} & \multicolumn{3}{c}{\textbf{GAM-PIG-One}} & \multicolumn{3}{c}{\textbf{GAM-NBG-One}} & \multicolumn{3}{c}{\textbf{GAM-NBIG-One}} & \multicolumn{3}{c}{\textbf{GLM-PG-One}} & \multicolumn{3}{c}{\textbf{GLM-PIG-One}} & \multicolumn{3}{c}{\textbf{GLM-NBG-One}} & \multicolumn{3}{c}{\textbf{GLM-NBIG-One}} \\ \hline \addlinespace[0.4ex]
		\textit{\texttt{Constant}} & -1.8783& (0.1185)&$^{***}$ & -1.8783& (0.1185)&$^{***}$ & -1.8788& (0.1197)&$^{***}$ & -1.8788& (0.1197)&$^{***}$ & -1.8666& (0.1151)&$^{***}$ & -1.8666& (0.1151)&$^{***}$ & -1.8670& (0.1161)&$^{***}$ & -1.8670& (0.1161)&$^{***}$ \\
		\textit{\texttt{ProductType}} & && & && & && & && & && & && & && & && \\
		\hspace{10pt}\textit{- Category 1} & -0.4659& (0.2019)&$^{*}$ & -0.4659& (0.2019)&$^{*}$ & -0.4631& (0.2032)&$^{*}$ & -0.4631& (0.2032)&$^{*}$ & -0.4672& (0.2018)&$^{*}$ & -0.4672& (0.2018)&$^{*}$ & -0.4643& (0.2032)&$^{*}$ & -0.4643& (0.2032)&$^{*}$ \\
		\hspace{10pt}\textit{- Category 2} & -0.3390& (0.0681)&$^{***}$ & -0.3390& (0.0681)&$^{***}$ & -0.3383& (0.0686)&$^{***}$ & -0.3383& (0.0686)&$^{***}$ & -0.3413& (0.0681)&$^{***}$ & -0.3413& (0.0681)&$^{***}$ & -0.3405& (0.0686)&$^{***}$ & -0.3405& (0.0686)&$^{***}$ \\
		\textit{\texttt{Year}} & && & && & && & && & && & && & && & && \\
		\hspace{10pt}\textit{- Category 1} & -0.3658& (0.0990)&$^{***}$ & -0.3658& (0.0990)&$^{***}$ & -0.3654& (0.0998)&$^{***}$ & -0.3654& (0.0998)&$^{***}$ & -0.3847& (0.0985)&$^{***}$ & -0.3847& (0.0985)&$^{***}$ & -0.3842& (0.0993)&$^{***}$ & -0.3842& (0.0993)&$^{***}$ \\
\end{tabular}}}
\end{table} %\newpage

\begin{table}[ht!]%[ht!]
    \caption{Parameter estimates and corresponding standard errors in parenthesis for travel insurance with dynamic univariate risk classification.}
    \label{T.C.9}\vspace{-3pt}
	\centerline{\scalebox{0.520}{\begin{tabular}{l r@{ }r@{}l r@{ }r@{}l r@{ }r@{}l r@{ }r@{}l r@{ }r@{}l r@{ }r@{}l r@{ }r@{}l r@{ }r@{}l }
	    \toprule \addlinespace[1ex] \vspace{1pt}
	    & \multicolumn{24}{c}{\textbf{Claim frequency}} \\
	    \cline{2-25} \addlinespace[0.4ex]
		\textbf{Risk factor} & \multicolumn{3}{c}{\textbf{GAM-PG-One}} & \multicolumn{3}{c}{\textbf{GAM-PIG-One}} & \multicolumn{3}{c}{\textbf{GAM-NBG-One}} & \multicolumn{3}{c}{\textbf{GAM-NBIG-One}} & \multicolumn{3}{c}{\textbf{GLM-PG-One}} & \multicolumn{3}{c}{\textbf{GLM-PIG-One}} & \multicolumn{3}{c}{\textbf{GLM-NBG-One}} & \multicolumn{3}{c}{\textbf{GLM-NBIG-One}} \\ \hline \addlinespace[0.4ex]
		\textit{\texttt{Constant}} & -2.9157& (0.0504)&$^{***}$ & -2.9157& (0.0504)&$^{***}$ & -2.9139& (0.0509)&$^{***}$ & -2.9139& (0.0509)&$^{***}$ & -3.0231& (0.0490)&$^{***}$ & -3.0231& (0.0490)&$^{***}$ & -3.0205& (0.0495)&$^{***}$ & -3.0205& (0.0495)&$^{***}$ \\
		\textit{\texttt{Region}} & && & && & && & && & && & && & && & && \\
		\hspace{10pt}\textit{- Category 1} & 0.3161& (0.0294)&$^{***}$ & 0.3161& (0.0294)&$^{***}$ & 0.3162& (0.0296)&$^{***}$ & 0.3162& (0.0296)&$^{***}$ & 0.3151& (0.0294)&$^{***}$ & 0.3151& (0.0294)&$^{***}$ & 0.3154& (0.0296)&$^{***}$ & 0.3154& (0.0296)&$^{***}$ \\
		\hspace{10pt}\textit{- Category 2} & 0.6887& (0.0265)&$^{***}$ & 0.6887& (0.0265)&$^{***}$ & 0.6905& (0.0268)&$^{***}$ & 0.6905& (0.0268)&$^{***}$ & 0.6820& (0.0265)&$^{***}$ & 0.6820& (0.0265)&$^{***}$ & 0.6840& (0.0268)&$^{***}$ & 0.6840& (0.0268)&$^{***}$ \\
		\textit{\texttt{Age}} & -0.0033& (0.0008)&$^{***}$ & -0.0033& (0.0008)&$^{***}$ & -0.0033& (0.0008)&$^{***}$ & -0.0033& (0.0008)&$^{***}$ & -0.0039& (0.0008)&$^{***}$ & -0.0039& (0.0008)&$^{***}$ & -0.0039& (0.0008)&$^{***}$ & -0.0039& (0.0008)&$^{***}$ \\
		\textit{\texttt{FamilySituation}} & && & && & && & && & && & && & && & && \\
\end{tabular}}}
\end{table} %\newpage

\begin{table}[ht!]%[ht!]
    \caption{Likelihood Ratio test statistics and corresponding p-values for the hypothesis that the alternative frequency model can better explain the in-sample data than the `null' frequency model, based on the asymptotic chi-squared distribution with $(k - 1) = 3$ and $1$ degrees of freedom for the one-product GAMs and GLMs, respectively.}
    \label{T.C.10}\vspace{-3pt}
	\centerline{\scalebox{0.645}{\begin{tabular}{l l r@{}l c c r@{}l c c r@{}l c c r@{}l c}
	    \toprule \addlinespace[1ex] \vspace{1pt}
	    &  & \multicolumn{3}{c}{\textbf{General liability}} &  & \multicolumn{3}{c}{\textbf{Home contents}} &  & \multicolumn{3}{c}{\textbf{Home}} &  & \multicolumn{3}{c}{\textbf{Travel}} \\
	    \cline{3-5} \cline{7-9} \cline{11-13} \cline{15-17} \addlinespace[0.4ex]
		\textbf{Alternative} & \textbf{Null} & \multicolumn{2}{c}{\textbf{Test statistic}} & \textbf{P-value} &  & \multicolumn{2}{c}{\textbf{Test statistic}} & \textbf{P-value} &  & \multicolumn{2}{c}{\textbf{Test statistic}} & \textbf{P-value} &  & \multicolumn{2}{c}{\textbf{Test statistic}} & \textbf{P-value} \\ \hline \addlinespace[0.4ex]
		\textit{GAM-PG-One} & GLM-P(I)G & 157.3024& & 0.0000 &  & 323.1985& & 0.0000 &  & 92.3563& & 0.0000 &  & 543.5951& & 0.0000 \\
		\textit{GAM-PIG-One} & GLM-P(I)G & 137.5704& & 0.0000 &  & 227.9287& & 0.0000 &  & 77.1978& & 0.0000 &  & 487.9059& & 0.0000 \\
		\textit{GAM-NBG-One} & GLM-NB(I)G & 173.9909& & 0.0000 &  & 408.9520& & 0.0000 &  & 105.5968& & 0.0000 &  & 589.5757& & 0.0000 \\
		\textit{GAM-NBIG-One} & GLM-NB(I)G & 154.2590& & 0.0000 &  & 313.6822& & 0.0000 &  & 90.4383& & 0.0000 &  & 533.8865& & 0.0000 \\
		\textit{GLM-PG-One} & GLM-P(I)G & 183.4555& & 0.0000 &  & 151.1394& & 0.0000 &  & 86.4301& & 0.0000 &  & 478.2224& & 0.0000 \\
		\textit{GLM-PIG-One} & GLM-P(I)G & 163.7235& & 0.0000 &  & 55.8697& & 0.0000 &  & 71.2715& & 0.0000 &  & 422.5331& & 0.0000 \\
		\textit{GLM-NBG-One} & GLM-NB(I)G & 200.2405& & 0.0000 &  & 243.8194& & 0.0000 &  & 99.8082& & 0.0000 &  & 526.1102& & 0.0000 \\
		\textit{GLM-NBIG-One} & GLM-NB(I)G & 184.7645& & 0.0000 &  & 148.5496& & 0.0000 &  & 84.6497& & 0.0000 &  & 470.4210& & 0.0000 \\
		\bottomrule
	\end{tabular}}}
\end{table} %\newpage

%%%%%%%%%%%%%%%%%% Dynamic multivariate %%%%%%%%%%%%%%%%
\begin{table}[ht!]%[ht!]
    \caption{Parameter estimates and corresponding standard errors in parenthesis for general liability insurance with dynamic multivariate risk classification.}
    \label{T.C.11}\vspace{-3pt}
	\centerline{\scalebox{0.495}{\begin{tabular}{l r@{ }r@{}l r@{ }r@{}l r@{ }r@{}l r@{ }r@{}l r@{ }r@{}l r@{ }r@{}l r@{ }r@{}l r@{ }r@{}l }
	    \toprule \addlinespace[1ex] \vspace{1pt}
	    & \multicolumn{24}{c}{\textbf{Claim frequency}} \\
	    \cline{2-25} \addlinespace[0.4ex]
		\textbf{Risk factor} & \multicolumn{3}{c}{\textbf{GAM-PG-Multi}} & \multicolumn{3}{c}{\textbf{GAM-PIG-Multi}} & \multicolumn{3}{c}{\textbf{GAM-NBG-Multi}} & \multicolumn{3}{c}{\textbf{GAM-NBIG-Multi}} & \multicolumn{3}{c}{\textbf{GLM-PG-Multi}} & \multicolumn{3}{c}{\textbf{GLM-PIG-Multi}} & \multicolumn{3}{c}{\textbf{GLM-NBG-Multi}} & \multicolumn{3}{c}{\textbf{GLM-NBIG-Multi}} \\ \hline \addlinespace[0.4ex]
		\textit{\texttt{Constant}} & -3.9201& (0.0575)&$^{***}$ & -3.9201& (0.0575)&$^{***}$ & -3.9209& (0.0580)&$^{***}$ & -3.9209& (0.0580)&$^{***}$ & -4.1074& (0.0440)&$^{***}$ & -4.1074& (0.0440)&$^{***}$ & -4.1053& (0.0443)&$^{***}$ & -4.0598& (0.0451)&$^{***}$ \\
		\textit{\texttt{FamilySituation}} & && & && & && & && & && & && & && & && \\
		\hspace{10pt}\textit{- Category 1} & 0.0355& (0.0541)& & 0.0355& (0.0541)& & 0.0359& (0.0544)& & 0.0359& (0.0544)& & 0.0747& (0.0537)& & 0.0747& (0.0537)& & 0.0745& (0.0540)& & 0.0700& (0.0540)& \\
		\hspace{10pt}\textit{- Category 2} & 0.9711& (0.0524)&$^{***}$ & 0.9711& (0.0524)&$^{***}$ & 0.9719& (0.0528)&$^{***}$ & 0.9719& (0.0528)&$^{***}$ & 0.9885& (0.0520)&$^{***}$ & 0.9885& (0.0520)&$^{***}$ & 0.9883& (0.0524)&$^{***}$ & 0.9855& (0.0524)&$^{***}$ \\
		\hspace{10pt}\textit{- Category 3} & 0.9340& (0.0696)&$^{***}$ & 0.9340& (0.0696)&$^{***}$ & 0.9333& (0.0703)&$^{***}$ & 0.9333& (0.0703)&$^{***}$ & 0.9576& (0.0694)&$^{***}$ & 0.9576& (0.0694)&$^{***}$ & 0.9578& (0.0702)&$^{***}$ & 0.9521& (0.0702)&$^{***}$ \\
		\hline \addlinespace[0.4ex]
		\textit{\texttt{General liability}} & && & && & && & && & && & && & && & && \\
\end{tabular}}}
\end{table} %\newpage

\begin{table}[ht!]%[ht!]
    \caption{Parameter estimates and corresponding standard errors in parenthesis for home contents insurance with dynamic multivariate risk classification.}
    \label{T.C.12}\vspace{-3pt}
	\centerline{\scalebox{0.455}{\begin{tabular}{l r@{ }r@{}l r@{ }r@{}l r@{ }r@{}l r@{ }r@{}l r@{ }r@{}l r@{ }r@{}l r@{ }r@{}l r@{ }r@{}l }
	    \toprule \addlinespace[1ex] \vspace{1pt}
	    & \multicolumn{24}{c}{\textbf{Claim frequency}} \\
	    \cline{2-25} \addlinespace[0.4ex]
		\textbf{Risk factor} & \multicolumn{3}{c}{\textbf{GAM-PG-Multi}} & \multicolumn{3}{c}{\textbf{GAM-PIG-Multi}} & \multicolumn{3}{c}{\textbf{GAM-NBG-Multi}} & \multicolumn{3}{c}{\textbf{GAM-NBIG-Multi}} & \multicolumn{3}{c}{\textbf{GLM-PG-Multi}} & \multicolumn{3}{c}{\textbf{GLM-PIG-Multi}} & \multicolumn{3}{c}{\textbf{GLM-NBG-Multi}} & \multicolumn{3}{c}{\textbf{GLM-NBIG-Multi}} \\ \hline \addlinespace[0.4ex]
		\textit{\texttt{Constant}} & -2.3683& (0.0745)&$^{***}$ & -2.3683& (0.0745)&$^{***}$ & -2.3669& (0.0756)&$^{***}$ & -2.3669& (0.0756)&$^{***}$ & -2.2199& (0.0718)&$^{***}$ & -2.2199& (0.0718)&$^{***}$ & -2.2162& (0.0729)&$^{***}$ & -2.2148& (0.0730)&$^{***}$ \\
		\textit{\texttt{ProductType}} & && & && & && & && & && & && & && & && \\
		\hspace{10pt}\textit{- Category 1} & -0.5660& (0.1504)&$^{***}$ & -0.5660& (0.1504)&$^{***}$ & -0.5675& (0.1519)&$^{***}$ & -0.5675& (0.1519)&$^{***}$ & -0.6382& (0.1498)&$^{***}$ & -0.6382& (0.1498)&$^{***}$ & -0.6397& (0.1513)&$^{***}$ & -0.6403& (0.1513)&$^{***}$ \\
		\hspace{10pt}\textit{- Category 2} & -0.1062& (0.0472)&$^{*}$ & -0.1062& (0.0472)&$^{*}$ & -0.1062& (0.0478)&$^{*}$ & -0.1062& (0.0478)&$^{*}$ & -0.1404& (0.0472)&$^{**}$ & -0.1404& (0.0472)&$^{**}$ & -0.1395& (0.0478)&$^{**}$ & -0.1395& (0.0478)&$^{**}$ \\
		\textit{\texttt{Year}} & && & && & && & && & && & && & && & && \\
		\hspace{10pt}\textit{- Category 1} & 0.0088& (0.0670)& & 0.0088& (0.0670)& & 0.0093& (0.0680)& & 0.0093& (0.0680)& & -0.0326& (0.0666)& & -0.0326& (0.0666)& & -0.0330& (0.0676)& & -0.0335& (0.0676)& \\
\end{tabular}}}
\end{table} %\newpage

\begin{table}[ht!]%[ht!]
    \caption{Parameter estimates and corresponding standard errors in parenthesis for home insurance with dynamic multivariate risk classification.}
    \label{T.C.13}\vspace{-3pt}
	\centerline{\scalebox{0.455}{\begin{tabular}{l r@{ }r@{}l r@{ }r@{}l r@{ }r@{}l r@{ }r@{}l r@{ }r@{}l r@{ }r@{}l r@{ }r@{}l r@{ }r@{}l }
	    \toprule \addlinespace[1ex] \vspace{1pt}
	    & \multicolumn{24}{c}{\textbf{Claim frequency}} \\
	    \cline{2-25} \addlinespace[0.4ex]
		\textbf{Risk factor} & \multicolumn{3}{c}{\textbf{GAM-PG-Multi}} & \multicolumn{3}{c}{\textbf{GAM-PIG-Multi}} & \multicolumn{3}{c}{\textbf{GAM-NBG-Multi}} & \multicolumn{3}{c}{\textbf{GAM-NBIG-Multi}} & \multicolumn{3}{c}{\textbf{GLM-PG-Multi}} & \multicolumn{3}{c}{\textbf{GLM-PIG-Multi}} & \multicolumn{3}{c}{\textbf{GLM-NBG-Multi}} & \multicolumn{3}{c}{\textbf{GLM-NBIG-Multi}} \\ \hline \addlinespace[0.4ex]
		\textit{\texttt{Constant}} & -2.0041& (0.1212)&$^{***}$ & -2.0041& (0.1212)&$^{***}$ & -2.0047& (0.1223)&$^{***}$ & -2.0047& (0.1223)&$^{***}$ & -1.8587& (0.1157)&$^{***}$ & -1.8587& (0.1157)&$^{***}$ & -1.8594& (0.1167)&$^{***}$ & -1.8610& (0.1168)&$^{***}$ \\
		\textit{\texttt{ProductType}} & && & && & && & && & && & && & && & && \\
		\hspace{10pt}\textit{- Category 1} & -0.4684& (0.2020)&$^{*}$ & -0.4684& (0.2020)&$^{*}$ & -0.4656& (0.2033)&$^{*}$ & -0.4656& (0.2033)&$^{*}$ & -0.4700& (0.2019)&$^{*}$ & -0.4700& (0.2019)&$^{*}$ & -0.4671& (0.2032)&$^{*}$ & -0.4669& (0.2032)&$^{*}$ \\
		\hspace{10pt}\textit{- Category 2} & -0.3454& (0.0683)&$^{***}$ & -0.3454& (0.0683)&$^{***}$ & -0.3446& (0.0688)&$^{***}$ & -0.3446& (0.0688)&$^{***}$ & -0.3453& (0.0682)&$^{***}$ & -0.3453& (0.0682)&$^{***}$ & -0.3445& (0.0687)&$^{***}$ & -0.3444& (0.0687)&$^{***}$ \\
		\textit{\texttt{Year}} & && & && & && & && & && & && & && & && \\
		\hspace{10pt}\textit{- Category 1} & -0.3234& (0.0994)&$^{**}$ & -0.3234& (0.0994)&$^{**}$ & -0.3228& (0.1002)&$^{**}$ & -0.3228& (0.1002)&$^{**}$ & -0.3853& (0.0986)&$^{***}$ & -0.3853& (0.0986)&$^{***}$ & -0.3848& (0.0994)&$^{***}$ & -0.3839& (0.0994)&$^{***}$ \\
\end{tabular}}}
\end{table} %\newpage

\begin{table}[ht!]%[ht!]
    \caption{Parameter estimates and corresponding standard errors in parenthesis for travel insurance with dynamic multivariate risk classification.}
    \label{T.C.14}\vspace{-3pt}
	\centerline{\scalebox{0.495}{\begin{tabular}{l r@{ }r@{}l r@{ }r@{}l r@{ }r@{}l r@{ }r@{}l r@{ }r@{}l r@{ }r@{}l r@{ }r@{}l r@{ }r@{}l }
	    \toprule \addlinespace[1ex] \vspace{1pt}
	    & \multicolumn{24}{c}{\textbf{Claim frequency}} \\
	    \cline{2-25} \addlinespace[0.4ex]
		\textbf{Risk factor} & \multicolumn{3}{c}{\textbf{GAM-PG-Multi}} & \multicolumn{3}{c}{\textbf{GAM-PIG-Multi}} & \multicolumn{3}{c}{\textbf{GAM-NBG-Multi}} & \multicolumn{3}{c}{\textbf{GAM-NBIG-Multi}} & \multicolumn{3}{c}{\textbf{GLM-PG-Multi}} & \multicolumn{3}{c}{\textbf{GLM-PIG-Multi}} & \multicolumn{3}{c}{\textbf{GLM-NBG-Multi}} & \multicolumn{3}{c}{\textbf{GLM-NBIG-Multi}} \\ \hline \addlinespace[0.4ex]
		\textit{\texttt{Constant}} & -2.9278& (0.0506)&$^{***}$ & -2.9278& (0.0506)&$^{***}$ & -2.9261& (0.0511)&$^{***}$ & -2.9261& (0.0511)&$^{***}$ & -3.0310& (0.0491)&$^{***}$ & -3.0310& (0.0491)&$^{***}$ & -3.0284& (0.0496)&$^{***}$ & -3.0283& (0.0496)&$^{***}$ \\
		\textit{\texttt{Region}} & && & && & && & && & && & && & && & && \\
		\hspace{10pt}\textit{- Category 1} & 0.3141& (0.0294)&$^{***}$ & 0.3141& (0.0294)&$^{***}$ & 0.3143& (0.0297)&$^{***}$ & 0.3143& (0.0297)&$^{***}$ & 0.3137& (0.0294)&$^{***}$ & 0.3137& (0.0294)&$^{***}$ & 0.3139& (0.0297)&$^{***}$ & 0.3139& (0.0297)&$^{***}$ \\
		\hspace{10pt}\textit{- Category 2} & 0.6879& (0.0265)&$^{***}$ & 0.6879& (0.0265)&$^{***}$ & 0.6898& (0.0268)&$^{***}$ & 0.6898& (0.0268)&$^{***}$ & 0.6836& (0.0265)&$^{***}$ & 0.6836& (0.0265)&$^{***}$ & 0.6856& (0.0268)&$^{***}$ & 0.6855& (0.0268)&$^{***}$ \\
		\textit{\texttt{Age}} & -0.0031& (0.0008)&$^{***}$ & -0.0031& (0.0008)&$^{***}$ & -0.0031& (0.0008)&$^{***}$ & -0.0031& (0.0008)&$^{***}$ & -0.0036& (0.0008)&$^{***}$ & -0.0036& (0.0008)&$^{***}$ & -0.0037& (0.0008)&$^{***}$ & -0.0037& (0.0008)&$^{***}$ \\
		\textit{\texttt{FamilySituation}} & && & && & && & && & && & && & && & && \\
\end{tabular}}}
\end{table} %\newpage

\begin{table}[ht!]%[ht!]
    \caption{Likelihood Ratio test statistics and corresponding p-values for the hypothesis that the alternative frequency model can better explain the in-sample data than the `null' frequency model, based on the asymptotic chi-squared distribution with $3(k - 1) = 9$ and $3$ degrees of freedom for the multi-product GAMs and GLMs, respectively.}
    \label{T.C.15}\vspace{-3pt}
	\centerline{\scalebox{0.625}{\begin{tabular}{l l r@{}l c c r@{}l c c r@{}l c c r@{}l c}
	    \toprule \addlinespace[1ex] \vspace{1pt}
	    &  & \multicolumn{3}{c}{\textbf{General liability}} &  & \multicolumn{3}{c}{\textbf{Home contents}} &  & \multicolumn{3}{c}{\textbf{Home}} &  & \multicolumn{3}{c}{\textbf{Travel}} \\
	    \cline{3-5} \cline{7-9} \cline{11-13} \cline{15-17} \addlinespace[0.4ex]
		\textbf{Alternative} & \textbf{Null} & \multicolumn{2}{c}{\textbf{Test statistic}} & \textbf{P-value} &  & \multicolumn{2}{c}{\textbf{Test statistic}} & \textbf{P-value} &  & \multicolumn{2}{c}{\textbf{Test statistic}} & \textbf{P-value} &  & \multicolumn{2}{c}{\textbf{Test statistic}} & \textbf{P-value} \\ \hline \addlinespace[0.4ex]
		\textit{GAM-PG-Multi} & GAM-PG-One & 59.7632& & 0.0000 &  & 230.7485& & 0.0000 &  & 78.1863& & 0.0000 &  & 51.3983& & 0.0000 \\
		\textit{GAM-PIG-Multi} & GAM-PIG-One & 59.7632& & 0.0000 &  & 230.7485& & 0.0000 &  & 78.1863& & 0.0000 &  & 51.3983& & 0.0000 \\
		\textit{GAM-NBG-Multi} & GAM-NBG-One & 58.5701& & 0.0000 &  & 222.4131& & 0.0000 &  & 76.9025& & 0.0000 &  & 50.5386& & 0.0000 \\
		\textit{GAM-NBIG-Multi} & GAM-NBIG-One & 58.5701& & 0.0000 &  & 222.4131& & 0.0000 &  & 76.9025& & 0.0000 &  & 50.5386& & 0.0000 \\
		\textit{GLM-PG-Multi} & GLM-PG-One & 3.9593& & 0.2659 &  & 95.3309& & 0.0000 &  & 1.9004& & 0.5933 &  & 6.5193& & 0.0889 \\
		\textit{GLM-PIG-Multi} & GLM-PIG-One & 3.9593& & 0.2659 &  & 95.3309& & 0.0000 &  & 1.9004& & 0.5933 &  & 6.5193& & 0.0889 \\
		\textit{GLM-NBG-Multi} & GLM-NBG-One & 4.0103& & 0.2604 &  & 90.7355& & 0.0000 &  & 1.8392& & 0.6064 &  & 6.2359& & 0.1007 \\
		\textit{GLM-NBIG-Multi} & GLM-NBIG-One & 5.0968& & 0.1648 &  & 87.9611& & 0.0000 &  & 1.5199& & 0.6777 &  & 5.6160& & 0.1319 \\
		\bottomrule
	\end{tabular}}}
\end{table} %\newpage

%%%%%%%%%%%%%%%%%% Dynamic piecewise linear multivariate %%%%%%%%%%%%%%%%
\begin{table}[ht!]%[ht!]
    \caption{Parameter estimates and corresponding standard errors in parenthesis for general liability insurance with dynamic multivariate risk classification, extended to piecewise linear specifications.}
    \label{T.C.16}\vspace{-3pt}
	\centerline{\scalebox{0.770}{\begin{tabular}{l r@{ }r@{}l r@{ }r@{}l r@{ }r@{}l r@{ }r@{}l }
	    \toprule \addlinespace[1ex] \vspace{1pt}
	    & \multicolumn{12}{c}{\textbf{Claim frequency}} \\
	    \cline{2-13} \addlinespace[0.4ex]
		\textbf{Risk factor} & \multicolumn{3}{c}{\textbf{GAM-PG-Multi-PL}} & \multicolumn{3}{c}{\textbf{GAM-PIG-Multi-PL}} & \multicolumn{3}{c}{\textbf{GAM-NBG-Multi-PL}} & \multicolumn{3}{c}{\textbf{GAM-NBIG-Multi-PL}} \\ \hline \addlinespace[0.4ex]
		\textit{\texttt{Constant}} & -3.9727& (0.0603)&$^{***}$ & -3.9727& (0.0603)&$^{***}$ & -3.9731& (0.0607)&$^{***}$ & -3.9731& (0.0607)&$^{***}$ \\
		\textit{\texttt{FamilySituation}} & && & && & && & && \\
		\hspace{10pt}\textit{- Category 1} & 0.0400& (0.0541)& & 0.0400& (0.0541)& & 0.0404& (0.0544)& & 0.0404& (0.0544)& \\
		\hspace{10pt}\textit{- Category 2} & 0.9744& (0.0524)&$^{***}$ & 0.9744& (0.0524)&$^{***}$ & 0.9752& (0.0528)&$^{***}$ & 0.9752& (0.0528)&$^{***}$ \\
		\hspace{10pt}\textit{- Category 3} & 0.9357& (0.0696)&$^{***}$ & 0.9357& (0.0696)&$^{***}$ & 0.9349& (0.0703)&$^{***}$ & 0.9349& (0.0703)&$^{***}$ \\
		\hline \addlinespace[0.4ex]
		\textit{\texttt{General liability}} & && & && & && & && \\
\end{tabular}}}
\end{table} %\newpage

\begin{table}[ht!]%[ht!]
    \caption{Parameter estimates and corresponding standard errors in parenthesis for home contents insurance with dynamic multivariate risk classification, extended to piecewise linear specifications.}
    \label{T.C.17}\vspace{-3pt}
	\centerline{\scalebox{0.770}{\begin{tabular}{l r@{ }r@{}l r@{ }r@{}l r@{ }r@{}l r@{ }r@{}l }
	    \toprule \addlinespace[1ex] \vspace{1pt}
	    & \multicolumn{12}{c}{\textbf{Claim frequency}} \\
	    \cline{2-13} \addlinespace[0.4ex]
		\textbf{Risk factor} & \multicolumn{3}{c}{\textbf{GAM-PG-Multi-PL}} & \multicolumn{3}{c}{\textbf{GAM-PIG-Multi-PL}} & \multicolumn{3}{c}{\textbf{GAM-NBG-Multi-PL}} & \multicolumn{3}{c}{\textbf{GAM-NBIG-Multi-PL}} \\ \hline \addlinespace[0.4ex]
		\textit{\texttt{Constant}} & -2.3329& (0.0751)&$^{***}$ & -2.3329& (0.0751)&$^{***}$ & -2.3317& (0.0762)&$^{***}$ & -2.3317& (0.0762)&$^{***}$ \\
		\textit{\texttt{ProductType}} & && & && & && & && \\
		\hspace{10pt}\textit{- Category 1} & -0.5848& (0.1504)&$^{***}$ & -0.5848& (0.1504)&$^{***}$ & -0.5863& (0.1519)&$^{***}$ & -0.5863& (0.1519)&$^{***}$ \\
		\hspace{10pt}\textit{- Category 2} & -0.1127& (0.0472)&$^{*}$ & -0.1127& (0.0472)&$^{*}$ & -0.1127& (0.0478)&$^{*}$ & -0.1127& (0.0478)&$^{*}$ \\
		\textit{\texttt{Year}} & && & && & && & && \\
		\hspace{10pt}\textit{- Category 1} & 0.0023& (0.0670)& & 0.0023& (0.0670)& & 0.0027& (0.0679)& & 0.0027& (0.0679)& \\
\end{tabular}}}
\end{table} %\newpage

\begin{table}[ht!]%[ht!]
    \caption{Parameter estimates and corresponding standard errors in parenthesis for home insurance with dynamic multivariate risk classification, extended to piecewise linear specifications.}
    \label{T.C.18}\vspace{-3pt}
	\centerline{\scalebox{0.721}{\begin{tabular}{l r@{ }r@{}l r@{ }r@{}l r@{ }r@{}l r@{ }r@{}l }
	    \toprule \addlinespace[1ex] \vspace{1pt}
	    & \multicolumn{12}{c}{\textbf{Claim frequency}} \\
	    \cline{2-13} \addlinespace[0.4ex]
		\textbf{Risk factor} & \multicolumn{3}{c}{\textbf{GAM-PG-Multi-PL}} & \multicolumn{3}{c}{\textbf{GAM-PIG-Multi-PL}} & \multicolumn{3}{c}{\textbf{GAM-NBG-Multi-PL}} & \multicolumn{3}{c}{\textbf{GAM-NBIG-Multi-PL}} \\ \hline \addlinespace[0.4ex]
		\textit{\texttt{Constant}} & -2.0188& (0.1220)&$^{***}$ & -2.0188& (0.1220)&$^{***}$ & -2.0192& (0.1231)&$^{***}$ & -2.0192& (0.1231)&$^{***}$ \\
		\textit{\texttt{ProductType}} & && & && & && & && \\
		\hspace{10pt}\textit{- Category 1} & -0.4685& (0.2020)&$^{*}$ & -0.4685& (0.2020)&$^{*}$ & -0.4657& (0.2033)&$^{*}$ & -0.4657& (0.2033)&$^{*}$ \\
		\hspace{10pt}\textit{- Category 2} & -0.3436& (0.0683)&$^{***}$ & -0.3436& (0.0683)&$^{***}$ & -0.3428& (0.0688)&$^{***}$ & -0.3428& (0.0688)&$^{***}$ \\
		\textit{\texttt{Year}} & && & && & && & && \\
		\hspace{10pt}\textit{- Category 1} & -0.3331& (0.0992)&$^{***}$ & -0.3331& (0.0992)&$^{***}$ & -0.3325& (0.1000)&$^{***}$ & -0.3325& (0.1000)&$^{***}$ \\
\end{tabular}}}
\end{table} %\newpage

\begin{table}[ht!]%[ht!]
    \caption{Parameter estimates and corresponding standard errors in parenthesis for travel insurance with dynamic multivariate risk classification, extended to piecewise linear specifications.}
    \label{T.C.19}\vspace{-3pt}
	\centerline{\scalebox{0.770}{\begin{tabular}{l r@{ }r@{}l r@{ }r@{}l r@{ }r@{}l r@{ }r@{}l }
	    \toprule \addlinespace[1ex] \vspace{1pt}
	    & \multicolumn{12}{c}{\textbf{Claim frequency}} \\
	    \cline{2-13} \addlinespace[0.4ex]
		\textbf{Risk factor} & \multicolumn{3}{c}{\textbf{GAM-PG-Multi-PL}} & \multicolumn{3}{c}{\textbf{GAM-PIG-Multi-PL}} & \multicolumn{3}{c}{\textbf{GAM-NBG-Multi-PL}} & \multicolumn{3}{c}{\textbf{GAM-NBIG-Multi-PL}} \\ \hline \addlinespace[0.4ex]
		\textit{\texttt{Constant}} & -2.9312& (0.0504)&$^{***}$ & -2.9312& (0.0504)&$^{***}$ & -2.9295& (0.0510)&$^{***}$ & -2.9295& (0.0510)&$^{***}$ \\
		\textit{\texttt{Region}} & && & && & && & && \\
		\hspace{10pt}\textit{- Category 1} & 0.3137& (0.0294)&$^{***}$ & 0.3137& (0.0294)&$^{***}$ & 0.3139& (0.0297)&$^{***}$ & 0.3139& (0.0297)&$^{***}$ \\
		\hspace{10pt}\textit{- Category 2} & 0.6871& (0.0265)&$^{***}$ & 0.6871& (0.0265)&$^{***}$ & 0.6890& (0.0268)&$^{***}$ & 0.6890& (0.0268)&$^{***}$ \\
		\textit{\texttt{Age}} & -0.0031& (0.0008)&$^{***}$ & -0.0031& (0.0008)&$^{***}$ & -0.0031& (0.0008)&$^{***}$ & -0.0031& (0.0008)&$^{***}$ \\
		\textit{\texttt{FamilySituation}} & && & && & && & && \\
\end{tabular}}}
\end{table} %\newpage % Appendix C Additional estimation results
}

\end{document}